\pdfoutput=0
\documentclass[11pt]{article}
\usepackage{axodraw}
\usepackage{epsfig}
\usepackage{amsfonts}
\usepackage{amsmath}
\usepackage{bbm,bm}
\usepackage{cite}
\allowdisplaybreaks[1]

\input pix.sty

\renewcommand{\vec}[1]{{\bf #1}}

\newcommand{\tinymsbar}{{\overline{\mbox{\tiny\rm{MS}}}}}
\newcommand{\Lambdamsbar}{{\Lambda_\tinymsbar}}

\newcommand{\EA}{E^\rmii{$(A)$}_i}
\newcommand{\EB}{E^\rmii{$(B)$}_i}
\newcommand{\BA}{\vec{v}\times\vec{B}^{ }_i} % {v\!\times\!B^\rmii{$(A)$}_i}

\newcommand{\BG}{\bar{A}}
\newcommand{\T}{\rmii{$T$}}
\newcommand{\Nf}{N_{\rm f}}
\newcommand{\Nc}{N_{\rm c}}

\newcommand{\E}{\rmii{$E$}}
\renewcommand{\B}{\rmii{$B$}}

\newcommand{\CF}{C_\rmii{F}}
\newcommand{\gB}{g_\rmii{B}}

\newcommand{\mD}{m_\rmii{D}}

\newcommand{\gammaE}{\gamma_\rmii{E}}

\newcommand{\rmO}{{\mathcal{O}}}
\newcommand{\bmu}{\bar\mu}
\newcommand{\CA}{C_\rmii{A}} % {\Nc}

\def\lsi{\raise0.3ex\hbox{$<$\kern-0.75em\raise-1.1ex\hbox{$\sim$}}}
\def\gsi{\raise0.3ex\hbox{$>$\kern-0.75em\raise-1.1ex\hbox{$\sim$}}}

\newcommand{\nF}{n_\rmii{F}}
\newcommand{\nB}{n_\rmii{B}}
\newcommand{\rmii}[1]{{\mbox{\tiny\rm{#1}}}}
\newcommand{\rmiii}[1]{{\mbox{\tiny{$\scriptstyle{\rm#1}$}}}}
\newcommand{\re}{\mathop{\mbox{Re}}}

\newcommand{\Tint}[1]{{\hbox{$\sum$}\!\!\!\!\!\!\!\int\,}_{\!\!\!\!\raise-0.9ex\hbox{$\scriptstyle{#1}$}}}
\newcommand{\Tinti}[1]{{{\Sigma}\!\!\!\!\raise0.3ex\hbox{$\int$}_\rmii{${#1}$}}}
\newcommand{\Tintip}[1]{{{\Sigma'}\!\!\!\!\!\raise0.3ex\hbox{$\int$}_\rmii{${#1}$}}}

\newcommand{\bi}{\begin{itemize}}
\newcommand{\ei}{\end{itemize}}
\newcommand{\hide}[1]{ }
\newcommand{\bsl}[1]{\,\slash\!\!\!\!{#1}\,}

\newcommand{\deltabar}{\raise-0.02em\hbox{$\bar{}$}\hspace*{-0.8mm}{\delta}}
\def\TAsc(#1,#2)(#3,#4,#5)%
{\SetWidth{2.0}\CArc(#1,#2)(#3,#4,#5)\SetWidth{1.0}}
\def\Lwidth{3}

\def\TAgl(#1,#2)(#3,#4,#5){\SetWidth{2.0}\PhotonArc(#1,#2)(#3,#4,#5){\Lwidth}%
{6.283 #3 mul 360 div #4 #5 sub #4 #5 sub mul sqrt mul Tdensity mul}%
\SetWidth{1.0}}
\def\TLgl(#1,#2)(#3,#4){\SetWidth{2.0}\Photon(#1,#2)(#3,#4){\Lwidth}
{#1 #3 sub #1 #3 sub mul #2 #4 sub #2 #4 sub mul add sqrt Tdensity mul}%
\SetWidth{1.0}}
\def\Lwidth{1.3}

\def\procQCDa{\picc{%
 \Line(0,10)(60,10)%
 \Line(28,8)(32,12)%
 \Line(28,12)(32,8)%
 \CBoxc(0,10)(4,4){Black}{White}
 \CBoxc(60,10)(4,4){Black}{White}
 \Gluon(16,10)(16,30){2}{3}
}}
\def\procQCDb{\picc{%
 \Line(0,10)(60,10)%
 \Line(28,8)(32,12)%
 \Line(28,12)(32,8)%
 \CBoxc(0,10)(4,4){Black}{White}
 \CBoxc(60,10)(4,4){Black}{White}
 \Gluon(44,10)(44,30){-2}{3}
}}
\def\procNRa{\picc{%
 \Line(0,10)(60,10)%
 \Line(28,8)(32,12)%
 \Line(28,12)(32,8)%
 \CBoxc(0,10)(4,4){Black}{White}
 \CBoxc(60,10)(4,4){Black}{White}
 \Gluon(31,10)(31,30){2}{3}
}}
\def\procQCDc{\picc{%
 \Line(0,10)(60,10)%
 \Line(28,8)(32,12)%
 \Line(28,12)(32,8)%
 \CBoxc(0,10)(4,4){Black}{White}
 \CBoxc(60,10)(4,4){Black}{White}
 \Lgl(16,10)(16,16)
 \Gluon(16,24)(16,32){2}{1}
 \CCirc(16,20){4}{Black}{Gray}
}}
\def\procQCDd{\picc{%
 \Line(0,10)(60,10)%
 \Line(28,8)(32,12)%
 \Line(28,12)(32,8)%
 \CBoxc(0,10)(4,4){Black}{White}
 \CBoxc(60,10)(4,4){Black}{White}
 \Lgl(44,10)(44,16)
 \Gluon(44,24)(44,32){-2}{1}
 \CCirc(44,20){4}{Black}{Gray}
}}
\def\procNRc{\picc{%
 \Line(0,10)(60,10)%
 \Line(28,8)(32,12)%
 \Line(28,12)(32,8)%
 \CBoxc(0,10)(4,4){Black}{White}
 \CBoxc(60,10)(4,4){Black}{White}
 \Lgl(31,10)(31,16)
 \Gluon(31,24)(31,32){2}{1}
 \CCirc(31,20){4}{Black}{Gray}
}}
\def\procQCDe{\picc{%
 \Line(0,10)(60,10)%
 \Line(28,8)(32,12)%
 \Line(28,12)(32,8)%
 \CBoxc(0,10)(4,4){Black}{White}
 \CBoxc(60,10)(4,4){Black}{White}
 \Photon(8,10)(15,22){1.5}{3}
 \Photon(22,10)(15,22){-1.5}{3}
 \Gluon(15,22)(15,30){2}{1}
}}
\def\procNRe{\picc{%
 \Line(0,10)(60,10)%
 \Line(28,8)(32,12)%
 \Line(28,12)(32,8)%
 \CBoxc(0,10)(4,4){Black}{White}
 \CBoxc(60,10)(4,4){Black}{White}
 \Photon(16,10)(23,22){1.5}{3}
 \Photon(30,10)(23,22){-1.5}{3}
 \Gluon(23,22)(23,30){2}{1}
}}
\def\procNRee{\picc{%
 \Line(0,10)(60,10)%
 \Line(28,8)(32,12)%
 \Line(28,12)(32,8)%
 \CBoxc(0,10)(4,4){Black}{White}
 \CBoxc(60,10)(4,4){Black}{White}
 \PhotonArc(16,16)(6,100,260){-1.5}{3.5}
 \PhotonArc(14,16)(6,-80,80){-1.5}{3.5}
 \Gluon(15,22)(15,30){2}{1}
}}
\def\procNReE{\picc{%
 \Line(0,10)(60,10)%
 \Line(28,8)(32,12)%
 \Line(28,12)(32,8)%
 \CBoxc(0,10)(4,4){Black}{White}
 \CBoxc(60,10)(4,4){Black}{White}
 \PhotonArc(31,16)(6,100,260){-1.5}{3.5}
 \PhotonArc(29,16)(6,-80,80){-1.5}{3.5}
 \Gluon(30,22)(30,30){2}{1}
}}
\def\procQCDf{\picc{%
 \Line(0,10)(60,10)%
 \Line(28,8)(32,12)%
 \Line(28,12)(32,8)%
 \CBoxc(0,10)(4,4){Black}{White}
 \CBoxc(60,10)(4,4){Black}{White}
 \Photon(23,10)(30,22){1.5}{3}
 \Photon(37,10)(30,22){-1.5}{3}
 \Gluon(30,22)(30,30){2}{1}
}}
\def\procQCDg{\picc{%
 \Line(0,10)(60,10)%
 \Line(28,8)(32,12)%
 \Line(28,12)(32,8)%
 \CBoxc(0,10)(4,4){Black}{White}
 \CBoxc(60,10)(4,4){Black}{White}
 \Photon(38,10)(45,22){1.5}{3}
 \Photon(52,10)(45,22){-1.5}{3}
 \Gluon(45,22)(45,30){2}{1}
}}
\def\procNRg{\picc{%
 \Line(0,10)(60,10)%
 \Line(28,8)(32,12)%
 \Line(28,12)(32,8)%
 \CBoxc(0,10)(4,4){Black}{White}
 \CBoxc(60,10)(4,4){Black}{White}
 \Photon(30,10)(37,22){1.5}{3}
 \Photon(44,10)(37,22){-1.5}{3}
 \Gluon(37,22)(37,30){2}{1}
}}
\def\procNRgg{\picc{%
 \Line(0,10)(60,10)%
 \Line(28,8)(32,12)%
 \Line(28,12)(32,8)%
 \CBoxc(0,10)(4,4){Black}{White}
 \CBoxc(60,10)(4,4){Black}{White}
 \PhotonArc(46,16)(6,100,260){-1.5}{3.5}
 \PhotonArc(44,16)(6,-80,80){-1.5}{3.5}
 \Gluon(45,22)(45,30){2}{1}
}}
\def\procQCDaA{\picc{%
 \Line(0,10)(60,10)%
 \Line(28,8)(32,12)%
 \Line(28,12)(32,8)%
 \CBoxc(0,10)(4,4){Black}{White}
 \CBoxc(60,10)(4,4){Black}{White}
 \Gluon(16,10)(16,30){2}{3}
 \PhotonArc(45,10)(8,181,359){1.3}{5.5}%
}}
\def\procNRaA{\picc{%
 \Line(0,10)(60,10)%
 \Line(28,8)(32,12)%
 \Line(28,12)(32,8)%
 \CBoxc(0,10)(4,4){Black}{White}
 \CBoxc(60,10)(4,4){Black}{White}
 \Gluon(16,10)(16,30){2}{3}
 \PhotonArc(45,5)(5,-270,90){1.3}{5.5}%
}}
\def\procNRaAA{\picc{%
 \Line(0,10)(60,10)%
 \Line(28,8)(32,12)%
 \Line(28,12)(32,8)%
 \CBoxc(0,10)(4,4){Black}{White}
 \CBoxc(60,10)(4,4){Black}{White}
 \Gluon(31,10)(31,30){2}{3}
 \PhotonArc(45,10)(8,181,359){1.3}{5.5}%
}}
\def\procNRaAB{\picc{%
 \Line(0,10)(60,10)%
 \Line(28,8)(32,12)%
 \Line(28,12)(32,8)%
 \CBoxc(0,10)(4,4){Black}{White}
 \CBoxc(60,10)(4,4){Black}{White}
 \Gluon(31,10)(31,30){2}{3}
 \PhotonArc(45,5)(5,-270,90){1.3}{5.5}%
}}
\def\procNRaAM{\picc{%
 \Line(0,10)(60,10)%
 \Line(39,8)(43,12)%
 \Line(39,12)(43,8)%
 \CBoxc(0,10)(4,4){Black}{White}
 \CBoxc(60,10)(4,4){Black}{White}
 \Gluon(16,10)(16,30){2}{3}
 \PhotonArc(32,10)(8,181,359){1.3}{5.5}%
}}
\def\procNRaAN{\picc{%
 \Line(0,10)(60,10)%
 \Line(39,8)(43,12)%
 \Line(39,12)(43,8)%
 \CBoxc(0,10)(4,4){Black}{White}
 \CBoxc(60,10)(4,4){Black}{White}
 \Gluon(32,10)(32,30){2}{3}
 \PhotonArc(32,10)(8,181,359){1.3}{5.5}%
}}
\def\procNRaAO{\picc{%
 \Line(0,10)(60,10)%
 \Line(29,8)(33,12)%
 \Line(29,12)(33,8)%
 \CBoxc(0,10)(4,4){Black}{White}
 \CBoxc(60,10)(4,4){Black}{White}
 \Gluon(44,10)(44,30){2}{3}
 \PhotonArc(22,10)(8,181,359){1.3}{5.5}%
}}
\def\procNRaAP{\picc{%
 \Line(0,10)(60,10)%
 \Line(21,8)(17,12)%
 \Line(21,12)(17,8)%
 \CBoxc(0,10)(4,4){Black}{White}
 \CBoxc(60,10)(4,4){Black}{White}
 \Gluon(44,10)(44,30){-2}{3}
 \PhotonArc(28,10)(8,181,359){1.3}{5.5}%
}}
\def\procNRaAQ{\picc{%
 \Line(0,10)(60,10)%
 \Line(21,8)(17,12)%
 \Line(21,12)(17,8)%
 \CBoxc(0,10)(4,4){Black}{White}
 \CBoxc(60,10)(4,4){Black}{White}
 \Gluon(28,10)(28,30){-2}{3}
 \PhotonArc(28,10)(8,181,359){1.3}{5.5}%
}}
\def\procNRaAR{\picc{%
 \Line(0,10)(60,10)%
 \Line(31,8)(27,12)%
 \Line(31,12)(27,8)%
 \CBoxc(0,10)(4,4){Black}{White}
 \CBoxc(60,10)(4,4){Black}{White}
 \Gluon(16,10)(16,30){-2}{3}
 \PhotonArc(38,10)(8,181,359){1.3}{5.5}%
}}
\def\procNRaAS{\picc{%
 \Line(0,10)(60,10)%
 \Line(39,8)(43,12)%
 \Line(39,12)(43,8)%
 \CBoxc(0,10)(4,4){Black}{White}
 \CBoxc(60,10)(4,4){Black}{White}
 \Gluon(24,10)(24,30){2}{3}
 \PhotonArc(32,10)(8,181,359){1.3}{5.5}%
}}
\def\procNRaASS{\picc{%
 \Line(0,10)(60,10)%
 \Line(22,8)(26,12)%
 \Line(22,12)(26,8)%
 \CBoxc(0,10)(4,4){Black}{White}
 \CBoxc(60,10)(4,4){Black}{White}
 \Gluon(24,10)(24,30){2}{3}
 \PhotonArc(32,10)(8,181,359){1.3}{5.5}%
}}
\def\procNRaAT{\picc{%
 \Line(0,10)(60,10)%
 \Line(21,8)(17,12)%
 \Line(21,12)(17,8)%
 \CBoxc(0,10)(4,4){Black}{White}
 \CBoxc(60,10)(4,4){Black}{White}
 \Gluon(36,10)(36,30){-2}{3}
 \PhotonArc(28,10)(8,181,359){1.3}{5.5}%
}}
\def\procNRaATT{\picc{%
 \Line(0,10)(60,10)%
 \Line(34,8)(38,12)%
 \Line(34,12)(38,8)%
 \CBoxc(0,10)(4,4){Black}{White}
 \CBoxc(60,10)(4,4){Black}{White}
 \Gluon(36,10)(36,30){-2}{3}
 \PhotonArc(28,10)(8,181,359){1.3}{5.5}%
}}
\def\procNRaAU{\picc{%
 \Line(0,10)(60,10)%
 \Line(40,8)(44,12)%
 \Line(40,12)(44,8)%
 \CBoxc(0,10)(4,4){Black}{White}
 \CBoxc(60,10)(4,4){Black}{White}
 \Gluon(16,10)(16,30){2}{3}
 \PhotonArc(24,10)(8,181,359){1.3}{5.5}%
}}
\def\procNRaAV{\picc{%
 \Line(0,10)(60,10)%
 \Line(28,8)(32,12)%
 \Line(28,12)(32,8)%
 \CBoxc(0,10)(4,4){Black}{White}
 \CBoxc(60,10)(4,4){Black}{White}
 \Gluon(22,10)(22,30){2}{3}
 \PhotonArc(30,10)(8,181,359){1.3}{5.5}%
}}
\def\procNRaAW{\picc{%
 \Line(0,10)(60,10)%
 \Line(20,8)(16,12)%
 \Line(20,12)(16,8)%
 \CBoxc(0,10)(4,4){Black}{White}
 \CBoxc(60,10)(4,4){Black}{White}
 \Gluon(28,10)(28,30){2}{3}
 \PhotonArc(36,10)(8,181,359){1.3}{5.5}%
}}
\def\procNRaAX{\picc{%
 \Line(0,10)(60,10)%
 \Line(20,8)(16,12)%
 \Line(20,12)(16,8)%
 \CBoxc(0,10)(4,4){Black}{White}
 \CBoxc(60,10)(4,4){Black}{White}
 \Gluon(44,10)(44,30){-2}{3}
 \PhotonArc(36,10)(8,181,359){1.3}{5.5}%
}}
\def\procNRaAY{\picc{%
 \Line(0,10)(60,10)%
 \Line(28,8)(32,12)%
 \Line(28,12)(32,8)%
 \CBoxc(0,10)(4,4){Black}{White}
 \CBoxc(60,10)(4,4){Black}{White}
 \Gluon(38,10)(38,30){-2}{3}
 \PhotonArc(30,10)(8,181,359){1.3}{5.5}%
}}
\def\procNRaAZ{\picc{%
 \Line(0,10)(60,10)%
 \Line(40,8)(44,12)%
 \Line(40,12)(44,8)%
 \CBoxc(0,10)(4,4){Black}{White}
 \CBoxc(60,10)(4,4){Black}{White}
 \Gluon(32,10)(32,30){-2}{3}
 \PhotonArc(24,10)(8,181,359){1.3}{5.5}%
}}
\def\procQCDaB{\picc{%
 \Line(0,10)(60,10)%
 \Line(28,8)(32,12)%
 \Line(28,12)(32,8)%
 \CBoxc(0,10)(4,4){Black}{White}
 \CBoxc(60,10)(4,4){Black}{White}
 \Gluon(16,10)(16,30){2}{3}
 \PhotonArc(30,10)(8,181,359){1.3}{5.5}%
}}
\def\procNRaBB{\picc{%
 \Line(0,10)(60,10)%
 \Line(28,8)(32,12)%
 \Line(28,12)(32,8)%
 \CBoxc(0,10)(4,4){Black}{White}
 \CBoxc(60,10)(4,4){Black}{White}
 \Gluon(31,10)(31,30){2}{3}
 \PhotonArc(30,10)(8,181,359){1.3}{5.5}%
}}
\def\procQCDaC{\picc{%
 \Line(0,10)(60,10)%
 \Line(33,8)(37,12)%
 \Line(33,12)(37,8)%
 \CBoxc(0,10)(4,4){Black}{White}
 \CBoxc(60,10)(4,4){Black}{White}
 \Gluon(22,10)(22,30){2}{3}
 \PhotonArc(30,30)(28,226,314){1.3}{10.5}%
}}
\def\procQCDaD{\picc{%
 \Line(0,10)(60,10)%
 \Line(42,8)(46,12)%
 \Line(42,12)(46,8)%
 \CBoxc(0,10)(4,4){Black}{White}
 \CBoxc(60,10)(4,4){Black}{White}
 \Gluon(16,10)(16,30){2}{3}
 \PhotonArc(30,10)(8,181,359){1.3}{5.5}%
}}
\def\procNRaD{\picc{%
 \Line(0,10)(60,10)%
 \Line(42,8)(46,12)%
 \Line(42,12)(46,8)%
 \CBoxc(0,10)(4,4){Black}{White}
 \CBoxc(60,10)(4,4){Black}{White}
 \Gluon(16,10)(16,30){2}{3}
 \PhotonArc(30,5)(5,-270,90){1.3}{5.5}%
}}
\def\procQCDaE{\picc{%
 \Line(0,10)(60,10)%
 \Line(28,8)(32,12)%
 \Line(28,12)(32,8)%
 \CBoxc(0,10)(4,4){Black}{White}
 \CBoxc(60,10)(4,4){Black}{White}
 \Gluon(16,10)(16,30){2}{3}
 \PhotonArc(15,10)(8,181,359){1.3}{5.5}%
}}
\def\procNRaEE{\picc{%
 \Line(0,10)(60,10)%
 \Line(28,8)(32,12)%
 \Line(28,12)(32,8)%
 \CBoxc(0,10)(4,4){Black}{White}
 \CBoxc(60,10)(4,4){Black}{White}
 \Gluon(31,10)(31,30){-2}{3}
 \PhotonArc(15,10)(8,181,359){1.3}{5.5}%
}}
\def\procNRaEF{\picc{%
 \Line(0,10)(60,10)%
 \Line(28,8)(32,12)%
 \Line(28,12)(32,8)%
 \CBoxc(0,10)(4,4){Black}{White}
 \CBoxc(60,10)(4,4){Black}{White}
 \Gluon(31,10)(31,30){-2}{3}
 \PhotonArc(15,5)(5,-270,90){1.3}{5.5}%
}}
\def\procQCDaF{\picc{%
 \Line(0,10)(60,10)%
 \Line(38,8)(42,12)%
 \Line(38,12)(42,8)%
 \CBoxc(0,10)(4,4){Black}{White}
 \CBoxc(60,10)(4,4){Black}{White}
 \Gluon(30,10)(30,30){2}{3}
 \PhotonArc(15,10)(8,181,359){1.3}{5.5}%
}}
\def\procNRaF{\picc{%
 \Line(0,10)(60,10)%
 \Line(38,8)(42,12)%
 \Line(38,12)(42,8)%
 \CBoxc(0,10)(4,4){Black}{White}
 \CBoxc(60,10)(4,4){Black}{White}
 \Gluon(30,10)(30,30){2}{3}
 \PhotonArc(15,5)(5,-270,90){1.3}{5.5}%
}}
\def\procQCDbA{\picc{%
 \Line(0,10)(60,10)%
 \Line(28,8)(32,12)%
 \Line(28,12)(32,8)%
 \CBoxc(0,10)(4,4){Black}{White}
 \CBoxc(60,10)(4,4){Black}{White}
 \Gluon(44,10)(44,30){-2}{3}
 \PhotonArc(15,10)(8,181,359){1.3}{5.5}%
}}
\def\procNRbA{\picc{%
 \Line(0,10)(60,10)%
 \Line(28,8)(32,12)%
 \Line(28,12)(32,8)%
 \CBoxc(0,10)(4,4){Black}{White}
 \CBoxc(60,10)(4,4){Black}{White}
 \Gluon(44,10)(44,30){-2}{3}
 \PhotonArc(15,5)(5,-270,90){1.3}{5.5}%
}}
\def\procQCDbB{\picc{%
 \Line(0,10)(60,10)%
 \Line(28,8)(32,12)%
 \Line(28,12)(32,8)%
 \CBoxc(0,10)(4,4){Black}{White}
 \CBoxc(60,10)(4,4){Black}{White}
 \Gluon(44,10)(44,30){-2}{3}
 \PhotonArc(30,10)(8,181,359){1.3}{5.5}%
}}
\def\procQCDbC{\picc{%
 \Line(0,10)(60,10)%
 \Line(23,8)(27,12)%
 \Line(23,12)(27,8)%
 \CBoxc(0,10)(4,4){Black}{White}
 \CBoxc(60,10)(4,4){Black}{White}
 \Gluon(38,10)(38,30){-2}{3}
 \PhotonArc(30,30)(28,226,314){1.3}{10.5}%
}}
\def\procQCDbD{\picc{%
 \Line(0,10)(60,10)%
 \Line(18,8)(14,12)%
 \Line(18,12)(14,8)%
 \CBoxc(0,10)(4,4){Black}{White}
 \CBoxc(60,10)(4,4){Black}{White}
 \Gluon(44,10)(44,30){-2}{3}
 \PhotonArc(30,10)(8,181,359){1.3}{5.5}%
}}
\def\procNRbD{\picc{%
 \Line(0,10)(60,10)%
 \Line(18,8)(14,12)%
 \Line(18,12)(14,8)%
 \CBoxc(0,10)(4,4){Black}{White}
 \CBoxc(60,10)(4,4){Black}{White}
 \Gluon(44,10)(44,30){-2}{3}
 \PhotonArc(30,5)(5,-270,90){1.3}{5.5}%
}}
\def\procQCDbE{\picc{%
 \Line(0,10)(60,10)%
 \Line(28,8)(32,12)%
 \Line(28,12)(32,8)%
 \CBoxc(0,10)(4,4){Black}{White}
 \CBoxc(60,10)(4,4){Black}{White}
 \Gluon(44,10)(44,30){-2}{3}
 \PhotonArc(45,10)(8,181,359){1.3}{5.5}%
}}
\def\procQCDbF{\picc{%
 \Line(0,10)(60,10)%
 \Line(22,8)(18,12)%
 \Line(22,12)(18,8)%
 \CBoxc(0,10)(4,4){Black}{White}
 \CBoxc(60,10)(4,4){Black}{White}
 \Gluon(30,10)(30,30){-2}{3}
 \PhotonArc(45,10)(8,181,359){1.3}{5.5}%
}}
\def\procNRbF{\picc{%
 \Line(0,10)(60,10)%
 \Line(22,8)(18,12)%
 \Line(22,12)(18,8)%
 \CBoxc(0,10)(4,4){Black}{White}
 \CBoxc(60,10)(4,4){Black}{White}
 \Gluon(30,10)(30,30){-2}{3}
 \PhotonArc(45,5)(5,-270,90){1.3}{5.5}%
}}
\def\procQCDaX{\picc{%
 \Line(0,10)(60,10)%
 \Line(28,8)(32,12)%
 \Line(28,12)(32,8)%
 \CBoxc(0,10)(4,4){Black}{White}
 \CBoxc(60,10)(4,4){Black}{White}
 \Gluon(16,10)(16,30){2}{3}
 \CCirc(45,10){2}{Black}{Black}
}}
\def\procNRaX{\picc{%
 \Line(0,10)(60,10)%
 \Line(28,8)(32,12)%
 \Line(28,12)(32,8)%
 \CBoxc(0,10)(4,4){Black}{White}
 \CBoxc(60,10)(4,4){Black}{White}
 \Gluon(31,10)(31,30){2}{3}
 \CCirc(45,10){2}{Black}{Black}
}}
\def\procQCDaY{\picc{%
 \Line(0,10)(60,10)%
 \Line(43,8)(47,12)%
 \Line(43,12)(47,8)%
 \CBoxc(0,10)(4,4){Black}{White}
 \CBoxc(60,10)(4,4){Black}{White}
 \Gluon(16,10)(16,30){2}{3}
 \CCirc(30,10){2}{Black}{Black}
}}
\def\procQCDaZ{\picc{%
 \Line(0,10)(60,10)%
 \Line(43,8)(47,12)%
 \Line(43,12)(47,8)%
 \CBoxc(0,10)(4,4){Black}{White}
 \CBoxc(60,10)(4,4){Black}{White}
 \Gluon(31,10)(31,30){2}{3}
 \CCirc(15,10){2}{Black}{Black}
}}
\def\procQCDbX{\picc{%
 \Line(0,10)(60,10)%
 \Line(28,8)(32,12)%
 \Line(28,12)(32,8)%
 \CBoxc(0,10)(4,4){Black}{White}
 \CBoxc(60,10)(4,4){Black}{White}
 \Gluon(44,10)(44,30){-2}{3}
 \CCirc(15,10){2}{Black}{Black}
}}
\def\procNRbX{\picc{%
 \Line(0,10)(60,10)%
 \Line(28,8)(32,12)%
 \Line(28,12)(32,8)%
 \CBoxc(0,10)(4,4){Black}{White}
 \CBoxc(60,10)(4,4){Black}{White}
 \Gluon(31,10)(31,30){-2}{3}
 \CCirc(15,10){2}{Black}{Black}
}}
\def\procQCDbY{\picc{%
 \Line(0,10)(60,10)%
 \Line(13,8)(17,12)%
 \Line(13,12)(17,8)%
 \CBoxc(0,10)(4,4){Black}{White}
 \CBoxc(60,10)(4,4){Black}{White}
 \Gluon(44,10)(44,30){-2}{3}
 \CCirc(30,10){2}{Black}{Black}
}}
\def\procQCDbZ{\picc{%
 \Line(0,10)(60,10)%
 \Line(13,8)(17,12)%
 \Line(13,12)(17,8)%
 \CBoxc(0,10)(4,4){Black}{White}
 \CBoxc(60,10)(4,4){Black}{White}
 \Gluon(29,10)(29,30){-2}{3}
 \CCirc(45,10){2}{Black}{Black}
}}
\makeatletter \@addtoreset{equation}{section} \makeatother

\makeatletter
\renewcommand\section{\@startsection {section}{1}{\z@}%
                                   {-5.5ex \@plus -1ex \@minus -.2ex}% bfr-
                                   {2.3ex \@plus.2ex}%
                                   {\normalfont\large\bfseries}}
\renewcommand\subsection{\@startsection{subsection}{2}{\z@}%
                                     {-3.25ex\@plus -1ex \@minus -.2ex}%
                                     {1.5ex \@plus .2ex}%
                                     {\normalfont\normalsize\bfseries}}
\renewcommand\thesection {\@arabic\c@section}
\renewcommand\thesubsection   {\thesection.\@arabic\c@subsection}
\renewcommand{\@seccntformat}[1]{%
\csname the#1\endcsname.\hspace{1.0em}}
\makeatother
\begin{document}

\flushbottom

%%%%%%%%%%%%%%%%%%%%%%%%%%% TITLE/COVER %%%%%%%%%%%%%%%%%%%%%%%%%%%%%%%%%

\begin{titlepage}

\begin{flushright}
% OUTLINE  \\ 
% DRAFT \\ 
% arXiv:2103.14270\\ 
June 2021
\end{flushright}
\begin{centering}
\vfill

{\Large{\bf
 1-loop matching of a thermal Lorentz force
}} 

\vspace{0.8cm}

M.~Laine % $^\rmi{a}$
 
\vspace{0.8cm}

% $^\rmi{a}$%
{\em
AEC, 
Institute for Theoretical Physics, 
University of Bern, \\ 
Sidlerstrasse 5, CH-3012 Bern, Switzerland \\}

\vspace*{0.8cm}

\mbox{\bf Abstract}
 
\end{centering}

\vspace*{0.3cm}
 
\noindent
Studying the diffusion and kinetic equilibration 
of heavy quarks within a hot QCD medium profits from the knowledge of 
a coloured Lorentz force that acts on them. Starting from the spatial
components of the vector current, and carrying out two matching
computations, one for the heavy quark mass scale ($M$) and another for
thermal scales ($\sqrt{MT}$, $T$), we determine 1-loop matching 
coefficients for
the electric and magnetic parts of a Lorentz force. The magnetic part
has a non-zero anomalous dimension, which agrees with that extracted
from two other considerations, one thermal and the other in vacuum. 
The matching coefficient could enable a lattice study of 
a colour-magnetic 2-point correlator.

\vfill

%% %\noindent
%% %PACS numbers: 
%% %11.10.Wx, %        Finite temperature field theory
%% %11.15.Ha, %        Lattice gauge theory  
%% %12.38.Bx, %        Perturbative calculations in QCD
%% %12.38.Mh, %        Quark--gluon plasma
%% %14.40.Nd, %        Bottom mesons
%% %\\
%% %Keywords: Thermal Field Theory, Heavy Quark Physics,
%%            Quark-Gluon Plasma, Lattice QCD
 
\end{titlepage}

\tableofcontents

%%%%%%%%%%%%%%%%%%%%%%%%%%% SECTION %%%%%%%%%%%%%%%%%%%%%%%%%%%%%%%%%%%%%%
%
\section{Introduction}

The motion of heavy probe particles is a classic tool
for extracting information about 
the microscopic properties of an interacting
statistical system. In heavy ion collision experiments, one 
manifestation of this philosophy is to inspect 
how efficiently heavy 
flavours (charm and bottom quarks) participate in hydrodynamic 
flow (cf.,\ e.g.,\ ref.~\cite{pheno}). 
In cosmology, assuming that dark matter is made of weakly interacting
massive particles, it would be important to know for how long they stay in
kinetic equilibrium with the other particles, as this may
affect, amongst others, structure formation
(cf.,\ e.g.,\ ref.~\cite{kinetic}). 

To be concrete, 
consider a particle whose mass~$M$ 
is much larger than the 
temperature~$T$. Given that the average (equilibrium) 
velocity  is below unity, $v^2 \sim 3 T/M \ll 1$,
and the (equilibrium) density 
is exponentially suppressed, 
$n \sim (\frac{MT}{2\pi})^{3/2} e^{-M/T }$, 
we find ourselves in a non-relativistic dilute regime. 
Thinking of a single such particle, and assuming that it carries
the gauge charge $g$, the classical Lorentz force acting on it reads
\be
 \frac{{\rm d} p^\mu_{ }}{{\rm d}t}
 = g F^{\mu\nu}_{ } v^{ }_\nu
 \;, \la{lorentz_em}
\ee
where $p^\mu_{ }$ is the four-momentum
and $v^\mu_{ } \equiv (1,\vec{v})$ is the velocity. 
The Lorentz force contains 
an electric part ($\sim g\vec{E}$) 
and 
a magnetic one ($\sim g\vec{v}\times\vec{B}$).
It has thus been argued that at
zeroth order in $\vec{v}$, heavy quarks are affected by
colour-electric forces~\cite{cst,kappaE}, whereas at first order
in $\vec{v}$, corrections originate from 
colour-magnetic ones~\cite{1overM}. 
For dark matter, we could similarly consider the forces
originating from the weak gauge group. 

Being a classical description, \eq\nr{lorentz_em} is guaranteed to 
hold only at large time scales where phase decoherence has taken place, 
$t \gg 1/(\alpha^2 T)$, where $\alpha = g^2/(4\pi)$. 
Due to their large inertia, the time scale associated with 
the kinetic equilibration of heavy particles 
is $\sim M / (\alpha^2 T^2)$~\cite{mt}.
For $M \gg T$, there should thus be a broad range of time scales
for which \eq\nr{lorentz_em} is valid. 
At the same time, thermal effects 
break Lorentz invariance and distinguish between electric
and magnetic fields, modifying the respective couplings
(cf.\ \eq\nr{F_IR}).
In fact, we recover an unmodified
\eq\nr{lorentz_em} only in vacuum,\footnote{%
 There is a famous history
 of quantum-mechanical derivations of the Lorentz force,
 cf.\ e.g.\ ref.~\cite{dyson}.
 }  
where the decoherence argument does not apply, 
but $M \gg \Lambdamsbar$ 
still provides for a hierarchy of time scales
(cf.\ \eq\nr{a_i_QCD_vac_nlo}).
 
Given that colour interactions are strong in QCD,  
their effects should be investigated up to the non-perturbative level. 
For  colour-electric forces, large-scale lattice simulations
have indeed been carried out 
in recent years~\cite{lat2,lat25,lat3,lat4,lat5,lat6}, 
whereas for the colour-magnetic
corrections, the challenge lies ahead of us. 
In preparation for this task, the goal of the current study is to 
clarify the renormalization of the colour-magnetic 
part of \eq\nr{lorentz_em}. 
 Specifically, we show how a divergence found in ref.~\cite{1overM}, 
 cf.\ \eq\nr{rhoB}, gets cancelled after the inclusion of the 
 proper matching coefficient.  

%%%%%%%%%%%%%%%%%%%%%%%%%%% SECTION %%%%%%%%%%%%%%%%%%%%%%%%%%%%%%%%%%%%%%
%
\section{Outline of a procedure}
\la{se:formulation}

Let us consider the vector current, 
$J^\rmii{QCD}_\mu = \bar\psi \gamma^{ }_\mu\psi$, 
associated with one heavy flavour in QCD.\footnote{% 
 We do not elaborate
 on the overall factors $\pm i, \pm 1$ of the various operators, 
 on one hand because these play no role in the end, 
 on the other because we work in Euclidean spacetime, 
 with Euclidean Dirac matrices, 
 and then additional factors may originate from the time coordinate,
 temporal gauge field components, 
 spatial Dirac matrices, 
 and raising/lowering of indices.
 It would be a distraction to discuss all of them.  
 } 
The spatial integral over 
the zeroth component, $\int_\vec{x} J^\rmii{QCD}_0$, 
measures the net number of this
species (particles minus antiparticles), and is conserved
in the absence of weak interactions. In contrast, 
the spatial components, $\int_\vec{x} J^\rmii{QCD}_i$, 
are not conserved. They
measure velocities, and velocities can be changed by elastic reactions. 

Following \eq\nr{lorentz_em}, our focus here is on time
derivatives of velocities, i.e.\ accelerations. The QCD 
operator that we are interested in can formally be expressed as 
$\partial^{ }_0 \int_\vec{x} J^\rmii{QCD}_i$. In a vacuum setting, we could
take matrix elements of this operator in the presence
of a background gauge field $\BG(Q)$~\cite{lfa}, 
where $Q = (q^{ }_0,\vec{q})$ is a four-momentum. 
As we are aiming at an infrared (IR) description, 
$Q$ is considered small compared with other energy scales.
Schematically, then, 
we could consider matrix elements like 
\ba
 \bigl\langle \vec{p}^{ }_1 \bigl| 
 \, 
  \bigl[ \partial^{ }_0 
  {\textstyle \int_\vec{x}} J^\rmii{QCD}_i \bigr]^{ }_{\BG(Q)}
 \, 
 \bigr| \vec{p}^{ }_2 \bigr\rangle 
 & \simeq &
 \delta^{(3)}_{ } (\vec{p}^{ }_2 + \vec{q} - \vec{p}^{ }_1 )
 \, \mathcal{A}^\rmii{QCD}_i[\BG(Q)] + \rmO(q_0^2,\vec{q}^2,\vec{v}^2)
 \;, \la{numerator_def1} \\ 
%%%
 \bigl\langle \vec{p}^{ }_1 \bigl| 
 \, 
   {\textstyle \int_\vec{x} J^\rmii{QCD}_0 }
 \, 
 \bigr| \vec{p}^{ }_2 \bigr\rangle 
 & \simeq &
 \delta^{(3)}_{ } (\vec{p}^{ }_2  - \vec{p}^{ }_1 )
 \, \mathcal{N}^\rmii{QCD}_0  + \rmO(\vec{v}^2)
 \;, \la{denominator_def1}
\ea
where the precise way to extract the external states will 
be discussed presently, and $\vec{v}$ is the heavy-quark 
velocity in the medium rest frame. 

The matrix elements in 
\eqs\nr{numerator_def1} and \nr{denominator_def1}
are subject to wave function renormalization,
which drops out in the ratio
\be
 a^\rmii{QCD}_i \; \equiv  \; \frac{\mathcal{A}^\rmii{QCD}_i}
                               {\mathcal{N}^\rmii{QCD}_0}
 \;. \la{ratio_def1}
\ee
It is for the cause of such an acceleration, 
multiplied by a (thermally corrected) pole mass~$M$, that we would 
like to find an operator reminiscent of the Lorentz force. 

Before proceeding, we note that for the thermal effects 
that we are mostly concerned with, the 
notion of matrix elements such as \eqs\nr{numerator_def1} and
\nr{denominator_def1} is ambiguous. 
Therefore, we generalize the definitions to certain ``partition functions'', 
defined in configuration space. 
Let the Euclidean time coordinate be $\tau$
and a generic spatially averaged operator $O(\tau)$. The time
direction is compact and is chosen to lie in the interval
$\tau\in(-\frac{\beta}{2},\frac{\beta}{2})$, 
where $\beta \equiv \frac{1}{T}$ is the inverse temperature. 
In this language, we may consider the 3-point correlator
\be
     \Bigl\langle \, \tr \, \Bigl\{ \, 
     {\textstyle \int^{ }_\vec{y}} \,
     \psi(\tfr{\beta}{2},\vec{y}) 
     e^{ - i \vec{p}_1 \cdot \vec{y}} \,      
     \bigl[ O(0) \bigr]^{ }_{\BG(Q)} \,
     {\textstyle \int^{ }_\vec{x}} \,
     \bar\psi(-\tfr{\beta}{2},\vec{x}) 
     e^{   i \vec{p}_2 \cdot \vec{x}} \, \Bigr\} 
     \, \Bigr\rangle^{ }_{\T,\rmi{c}} 
 \;, \la{partition}
\ee
where $\langle ... \rangle^{ }_{\T,\rmi{c}}$ is a thermal average,
and $c$ stands for connected contractions. 
We take a trace in Dirac space, given 
that the operator we are interested in, 
cf.\ \eq\nr{F_IR}, is spin-independent. 
The part of this correlator 
proportional to $e^{-\beta M}$ originates from the single
heavy quark sector of the Hilbert space, and
gives the effects that we are interested in. 
In a vacuum setting, we may replace 
$\beta/2 \to +\infty$
and 
$-\beta/2 \to -\infty$. 
The leading asymptotics
picks up the desired states in this case, and matrix elements
analogous to \eqs\nr{numerator_def1}, \nr{denominator_def1}
are obtained as coefficients of the exponential fall-off, 
up to overall factors that drop out in \eq\nr{ratio_def1}. 

Let now $\theta$ represent a non-relativistic $2\Nc^{ }$-component
spinor, defined in the sense of Heavy Quark Effective Theory (HQET)
(cf.,\ e.g.,\ refs.~\cite{hqet,hqetprime,hqetlat} 
and references therein). 
This brings in {\em two} new ways to define the acceleration. 
The first is that we consider components of the Noether 
current, which now read 
$
 J^\rmii{HQET}_0 = \theta^\dagger\theta
$, 
$
 J^\rmii{HQET}_i = 
 - \theta^\dagger 
 ( i \overleftrightarrow{D}^{ }_i) \theta /(2M)
 + \rmO(1/M^2)
$,
and then compute matrix elements of 
$
 \partial^{ }_0 \int_\vec{x} J^\rmii{HQET}_i
$
and 
$
 \int_\vec{x} J^\rmii{HQET}_0
$, 
just like 
in \eqs\nr{numerator_def1} and \nr{denominator_def1}.  
% Even if the physical meaning of 
% $
%  \int_\vec{x} J^\rmii{HQET}_0
% $
% differs from that in QCD, 
% as the antiparticle part is not present, this has no influence
% in the single heavy quark sector of the Hilbert space. 

However, one can envisage a more radical reduction, 
to which we refer as an infrared (IR) description. 
This involves an operator reminiscent of the Lorentz force 
in \eq\nr{lorentz_em}, 
\be
 F^\rmii{IR}_i \; \equiv \; 
 - \, i \gB^{ }\,
 \theta^\dagger \bigl\{
                 Z^{ }_{\!\E} \, F^{ }_{i0} V^{ }_0 
              +  Z^{ }_{\!\B} \, F^{ }_{ij} V^{ }_j  
                \bigr\}
 \,\theta
 \;, \la{F_IR}
\ee
where 
$\gB^{ }$ denotes the bare gauge coupling, 
$
 -i \gB^{ } F^{ }_{\mu\nu} \equiv [D^{ }_\mu,D^{ }_\nu]
$
is a field strength, 
and $V = (i,\vec{v})$
is the (Euclidean) heavy-quark velocity.
It is important to stress that in the static picture of \eq\nr{F_IR}, 
the velocity $\vec{v}$ appears as an ``external'' parameter,
whose thermal distribution is fixed later on 
from separate considerations (cf.\ \se\ref{se:discussion}).

Defining matrix elements on the IR side as 
\ba
  \bigl\langle \vec{p}^{ }_1 \bigl| 
 \, 
  \bigl[ {\textstyle \int_\vec{x}} F^\rmii{IR}_i \bigr]^{ }_{\BG(Q)}
 \, 
 \bigr| \vec{p}^{ }_2 \bigr\rangle 
 & \simeq &
 \delta^{(3)}_{ } (\vec{p}^{ }_2 + \vec{q} - \vec{p}^{ }_1 )
 \, \mathcal{F}^\rmii{IR}_i[\BG(Q)]
 + \rmO(q_0^2,\vec{q}^2,\vec{v}^2)
 \;, \la{numerator_def3}  \\
%%%%%%%%%%%
 \bigl\langle \vec{p}^{ }_1 \bigl| 
 \, 
   {\textstyle \int_\vec{x}} J^\rmii{HQET}_0 
 \, 
 \bigr| \vec{p}^{ }_2 \bigr\rangle 
 & \simeq &
 \delta^{(3)}_{ } (\vec{p}^{ }_2  - \vec{p}^{ }_1 )
 \, \mathcal{N}^\rmii{IR}_0  + \rmO(\vec{v}^2)
 \;, \la{denominator_def3}
\ea
the goal would be to find matching coefficients 
$ 
 Z^{ }_{\!\E,\B}
$
such that (up to possible signature issues)
\be
 \frac{M \mathcal{A}^\rmiii{QCD}_i}{\mathcal{N}^\rmiii{QCD}_0}
 = 
 \frac{ \mathcal{F}^\rmiii{IR}_i}{\mathcal{N}^\rmiii{IR}_0}
 % + \rmO\bigl(\vec{v}^2\bigr)
 \;. \la{matching} 
\ee
This establishes the principal viability of
a dynamics like that in \eq\nr{lorentz_em}.\footnote{% 
 As alluded to above and demonstrated explicitly 
 in the later sections, the cancellation of singular
 propagators from the numerator and denominator separately
 requires the inclusion of thermal corrections
 in the pole mass $M$, however these effects are power-suppressed
 by $g^3 T/M$ or $g^2 T^2/M^2$, and in fact irrelevant for the definition
 of $Z^{ }_{\!\E,\B}$, which comprise of corrections only suppressed by $g^2$. 
 } 
Such dynamics has already been employed for deriving 
purely gluonic 2-point imaginary-time correlators,  
permitting to study features of heavy quark diffusion 
and kinetic equilibration~\cite{cst,kappaE,1overM}.

%%%%%%%%%%%%%%%%%%%%%%%%%%% SECTION %%%%%%%%%%%%%%%%%%%%%%%%%%%%%%%%%%%%%%
%
\section{QCD vacuum contribution}
\la{se:qcd_vacuum}

The purpose of the present section is to see how 
the objects of \eqs\nr{numerator_def1}--\nr{ratio_def1} look like 
at 1-loop level in vacuum QCD. Physically 
speaking, this amounts to accounting for the heavy quark mass scale, $M$. 
Even if the result will be quite simple
(cf.\ \eq\nr{a_i_QCD_vac_nlo}), we hope that a detailed exposition
can set the technical stage for the subsequent sections. 
The inverse of a heavy quark propagator is denoted by 
\be 
 \Delta^{ }_{P} \;\equiv\; P^2 + M^2 
 \;, \la{Delta}
\ee 
and $P$ normally denotes an on-shell four-momentum, 
i.e.\ $P^2 = -M^2$.

To get going, we evaluate the 3-point correlator 
of \eq\nr{partition} at leading order (LO), 
with the sink and source placed at
$\beta/2 \to y^{ }_0 $ 
and 
$-\beta/2 \to x^{ }_0 $, 
respectively.
For the denominator, the operator reads 
$
 O(0) = \int_\vec{x} \bar\psi \gamma^{ }_0 \psi
$.
The Wick contractions yield
\be
 \delta^{(3)}_{ }(\vec{p}^{ }_1 - \vec{p}^{ }_2)
 \int_{\omega^{ }_1,\omega^{ }_2}
 e^{i(\omega^{ }_1 y^{ }_0 - \omega^{ }_2 x^{ }_0)}
 \frac{
    \tr [ (-i \bsl{P}^{ }_{\!\!1} + M)
           \gamma^{ }_0 
           (-i \bsl{P}^{ }_{\!\!2} + M)]
      }{(\omega_1^2 + \epsilon_{p_1}^2) (\omega_2^2 + \epsilon_{p_2}^2) }
 \;, 
\ee
where
$
 \epsilon^{ }_p \equiv \sqrt{p^2 + M^2}
$
and 
$
 \int^{ }_{\omega_i} \equiv \int_{-\infty}^{\infty} 
 {\rm d}\omega^{ }_i / (2\pi) 
$.
Sending
$ 
 y^{ }_0\to +\infty
$,
$
 x^{ }_0\to -\infty
$, 
the integrals over $\omega^{ }_{1,2}$ pick up the poles 
at 
\be
 \omega^{ }_1 = i \epsilon^{ }_{p_1}
 \;, \quad
 \omega^{ }_2 = i \epsilon^{ }_{p_2}
 \;, \la{on-shell}
\ee 
respectively. 
As momentum conservation sets the two momenta equal, 
we denote $\vec{p} \; \equiv \; \vec{p}^{ }_1 = \vec{p}^{ }_2$.
The asymptotic wave functions 
$
 e^{-\epsilon^{ }_p y^{ }_0}
 \times 
 e^{\epsilon^{ }_p x^{ }_0}
$
are factored out, and this defines what we mean by 
the remaining matrix element. 
Taking the trace and expanding to leading order 
in~$
 \vec{v} \; \equiv \; {\vec{p}} / {\epsilon^{ }_p}
$,  
in accordance with \eq\nr{denominator_def1},  
we then obtain
\be
 \mathcal{N}_0^\rmii{QCD,vac} 
 = 2   + \rmO(\gB^2)
 \;,
 \la{N_0_QCD_vac_lo}
\ee
multiplied by a unit matrix in colour space that is suppressed from 
the notation. 

%%%%%%%%%%%%%%%%%%%%%%%%% FIGURE %%%%%%%%%%%%%%%%%%%%%%%%%%%%%%%%%%%%%%%%%
%
\begin{figure}[t]
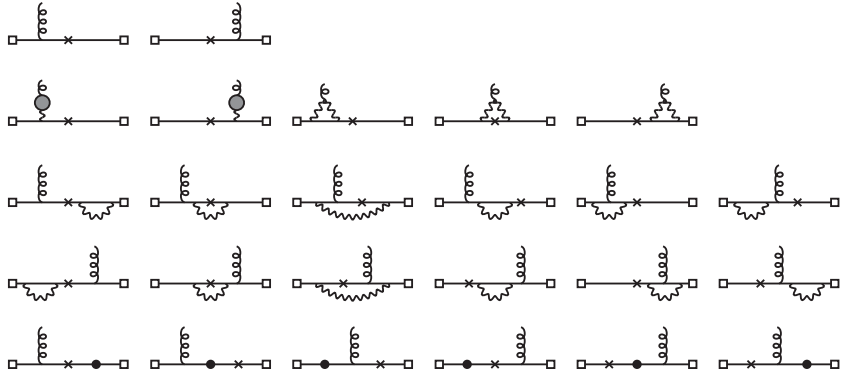


\hspace*{0.5cm}%
\begin{minipage}[c]{15cm}
\begin{eqnarray*}
 &&
 \hspace*{-1.0cm}
%%%%%%%%%%
 \procQCDa
 \hspace*{0.20cm}
 \procQCDb
%%%%%%%%%%
 \nn[2mm] 
 &&
 \hspace*{-1.0cm}
 \procQCDc
 \hspace*{0.20cm}
 \procQCDd
 \hspace*{0.20cm}
 \procQCDe
 \hspace*{0.20cm}
 \procQCDf
 \hspace*{0.20cm}
 \procQCDg
%%%%%%%%%%
 \nn[2mm] 
 &&
 \hspace*{-1.0cm}
 \procQCDaA
 \hspace*{0.20cm}
 \procQCDaB
 \hspace*{0.20cm}
 \procQCDaC
 \hspace*{0.20cm}
 \procQCDaD
 \hspace*{0.20cm}
 \procQCDaE
 \hspace*{0.20cm}
 \procQCDaF
%%%%%%%%%%
 \nn[2mm] 
 &&
 \hspace*{-1.0cm}
 \procQCDbA
 \hspace*{0.20cm}
 \procQCDbB
 \hspace*{0.20cm}
 \procQCDbC
 \hspace*{0.20cm}
 \procQCDbD
 \hspace*{0.20cm}
 \procQCDbE
 \hspace*{0.20cm}
 \procQCDbF
%%%%%%%%%%
 \nn[2mm] 
 &&
 \hspace*{-1.0cm}
 \procQCDaX
 \hspace*{0.20cm}
 \procQCDaY
 \hspace*{0.20cm}
 \procQCDaZ
 \hspace*{0.20cm}
 \procQCDbX
 \hspace*{0.20cm}
 \procQCDbY
 \hspace*{0.20cm}
 \procQCDbZ
% \nn[-10mm]
\end{eqnarray*}
\end{minipage}

\caption[a]{\small 
 The LO and NLO graphs contributing to the 3-point correlator 
 in full QCD. A solid line denotes a heavy quark, 
 a curly line an external gauge field, 
 a wavy line a dynamical gauge field, 
 a grey blob a 1-loop gauge field self-energy, 
 a solid circle a mass counterterm, 
 open squares a source and a sink,  
 and a cross the operator related to the conserved current
 or its time derivative.} 
 \la{fig:QCD}
\end{figure}
%
%%%%%%%%%%%%%%%%%%%%%%%%%%%%%%%%%%%%%%%%%%%%%%%%%%%%%%%%%%%%%%%%%%%%%%%%%%

Proceeding to the numerator, 
the operator can be expressed in momentum space as
\be
 \partial^{ }_0 \int_\vec{x} J^\rmii{QCD}_i 
 = 
 \int^{ }_{P^{ }_3,P^{ }_4}
 i (\omega^{ }_4 - \omega^{ }_3)\,
 \bar\psi(P^{ }_3) \gamma^{ }_i \psi(P^{ }_4)\,
 \delta^{(3)}_{ }(\vec{p}^{ }_4 - \vec{p}^{ }_3) 
 \;, \quad
 P^{ }_i = (\omega^{ }_i,\vec{p}^{ }_i)
 \;. \la{operator_QCD}
\ee
The diagrams to be computed are shown 
on the first row of \fig\ref{fig:QCD}. 
The key feature is that, after contracting the momenta to the 
external ones, i.e.\ $P^{ }_1$ and $P^{ }_2$, 
the prefactor 
$
 \omega^{ }_4 - \omega^{ }_3 = q^{ }_0 + \omega^{ }_2 - \omega^{ }_1
$
in \eq\nr{operator_QCD} is of $\rmO(Q)$, 
but there is an internal propagator (between the external gauge field
and the operator) which is of $\rmO(1/Q)$. 
These leading singularities cancel, leaving 
over terms of $\rmO(1)$: 
\be
 \frac{q^{ }_0 + \omega^{ }_2 - \omega^{ }_1}
      {\Delta^{ }_{P^{ }_1 - Q}}
 = \frac{1}{q^{ }_0 - i \epsilon^{ }_{p_1} - i \epsilon^{ }_{p_2}}
 \;, \quad
 \frac{q^{ }_0 + \omega^{ }_2 - \omega^{ }_1}
      {\Delta^{ }_{P^{ }_2 + Q}}
 = \frac{1}{q^{ }_0 + i \epsilon^{ }_{p_1} + i \epsilon^{ }_{p_2} }
 \;. \la{cancel_1}
\ee 
Here we made use of the overall momentum constraint
$
 \vec{p}^{ }_1 = \vec{p}^{ }_2 + \vec{q} 
$
and put the states on-shell according to \eq\nr{on-shell}. 
Subsequently we can insert 
\be
 \vec{p}^{ }_1 = \vec{p} + \frac{\vec{q}}{2}
 \;, \quad
 \vec{p}^{ }_2 = \vec{p} - \frac{\vec{q}}{2}
 \;, \quad
 \epsilon^{ }_{p_1} \approx \epsilon^{ }_p + \frac{\vec{v}\cdot\vec{q}}{2}
 \;, \quad
 \epsilon^{ }_{p_2} \approx \epsilon^{ }_p - \frac{\vec{v}\cdot\vec{q}}{2}
 \;, \la{taylor} 
\ee
and Taylor-expand to first order in $q^{ }_0$, $\vec{q}$ and $\vec{v}$.
For future convenience, we split electric fields into two parts, 
introducing
(while being again unconventional 
about factors of $i$)\hspace*{0.1mm}\footnote{% 
 As depicted in \figs\ref{fig:QCD}--\ref{fig:IR}, 
 we compute to linear order in the external gauge field 
 $\bar{A}^{ }_\mu \equiv \bar{A}^a_\mu T^a_{ }$, 
 whereby only the Abelian part appears in the external field strength.
 Here $T^a_{ }$ are Hermitean generators of SU($\Nc^{ }$). 
 } 
\be
 \EA \; \equiv \; i q^{ }_i \BG^{ }_0
 \;, \quad
 \EB \; \equiv \; i q^{ }_0 \BG^{ }_i
 \;, \quad
 \BA \; \equiv \; q^{ }_i \vec{v}\cdot\vec{\BG}
% \;, \quad \BB \; \equiv \;
                - \BG^{ }_i \vec{v}\cdot\vec{q} 
 \;. \la{fields}
\ee
Factoring out the same wave functions as above \eq\nr{N_0_QCD_vac_lo}, 
this leads to  
\be
 M\, \mathcal{A}_i^\rmii{QCD,vac}
 = 
 - \, 2 i \gB^{ } 
 \bigl[ 
   \EA - \EB + \BA % - \BB
 \bigr] 
 + \rmO(\gB^3)
 \;. \la{A_i_QCD_vac_lo}
\ee
Up to overall signature, 
the ratio of \eqs\nr{A_i_QCD_vac_lo} and \nr{N_0_QCD_vac_lo} 
yields a Lorentz force like in \eq\nr{lorentz_em}.

The task then is to proceed to next-to-leading order (NLO). 
For $\mathcal{N}^{ }_0$, the computation is relatively straightforward.
% (the diagrams are as in \fig\ref{fig:QCD} but disregarding the external
% gauge field). 
To remain consistently within the perturbative expansion, 
the 1-loop correction is evaluated at the location of the tree-level
poles, i.e.\ terms proportional to $P_1^2 + M^2$ or $P_2^2 + M^2$ 
are omitted. 
Here we simply state the result,  
\be
\frac{ \mathcal{N}^\rmii{QCD,vac}_0 % |^{ }_{\rmO(\gB^2)}
     }
     { 2  } 
 =  1 - \gB^2 \CF \int_R \biggl[ 
      \frac{4M^2}{R^2\Delta^2_{P-R}}
    + \biggl( 
          \frac{2}{\Delta^{ }_{P-R}} - \frac{D-2}{2M^2}
      \biggr)
      \biggl( 
          \frac{1}{\Delta^{ }_{P-R}} - \frac{1}{R^2}
      \biggr)
 \biggr]
 + \rmO(\gB^4)
 \;, \la{N_0_QCD_vac_nlo}
\ee
where $\CF \equiv (\Nc^2 - 1)/(2\Nc)$, 
$D=4-2\epsilon$ is the dimension of spacetime, 
$R$ is a gluon four-momentum, and $P$ is 
an on-shell heavy quark four-momentum, with $P^2 = - M^2$. 
Noting that scaleless integrals
vanish in dimensional regularization, and 
inserting non-vanishing master integrals from 
\eq\nr{vac_master_relation}, 
the explicit expression reads 
\be
\frac{ \mathcal{N}^\rmii{QCD,vac}_0  }
     { 2 } 
 = 1 + \frac{\gB^2 \CF\,\mu^{-2\epsilon} }{(4\pi)^2}
   \biggl( \frac{1}{\epsilon} + \ln\frac{\bmu^2}{M^2} + 4 \biggr) 
 + \rmO(\epsilon\,\gB^2,\gB^4)
 \;, \la{N_0_QCD_vac_nlo_expl} 
\ee
where $\bmu^2 \equiv 4\pi \mu^2 e^{-\gammaE}$ is the 
scale parameter of the $\msbar$ scheme. 

For the NLO computation of the numerator, 
let us give some more details. 
The diagrams are shown in \fig\ref{fig:QCD}. 
Actually, 
the gluon self-energy diagram is not needed, as it contains scaleless
integrals after the Taylor expansion in $Q$, 
and therefore vanishes in dimensional
regularization. The only exception is the loop
containing the heavy quark itself. The effect from here amounts
to the contribution that the heavy quark gives to the running of the 
gauge coupling. As the low-energy side of our matching is a theory
without the heavy quark, and we normally refer to the gauge coupling
of that theory, this effect is trivially included. 

Carrying out colour contractions
in the other diagrams, there are two parts, one proportional
to $\CF$ and the other to $\CA \equiv \Nc$. The part proportional to $\CF$ is 
quite IR sensitive: whereas at leading order there is one propagator
of $\rmO(1/Q)$, now there are two such propagators. These poles cancel
only by working in the pole mass scheme, whereby the mass counterterm
is chosen as ($M_\rmii{B}^2 = M^2 + \delta M^2$) 
\be 
 \delta M^2 = -\gB^2\CF \int_R
 \biggl[ 
   \frac{4 M^2}{R^2 \Delta^{ }_{P-R}}
  + 
   (D-2)\biggl( \frac{1}{R^2} - \frac{1}{\Delta^{ }_{P-R}} \biggr)
 \biggr] 
 + \rmO(\gB^4)
 \;. \la{deltaM_QCD}
\ee

We denote again the on-shell momenta of the external legs by 
$P^{ }_1$ and $P^{ }_2$, with $P^{ }_i = (\omega^{ }_i,\vec{p}^{ }_i)$; 
the dynamical gluon momentum by $R$; 
and the external gluon momentum by $Q$. After taking the Dirac trace, 
the first step is to eliminate scalar products like $R\cdot P^{ }_i$
or $Q\cdot R$, by completing squares and cancelling against denominators.
The key issue is to verify that, after including the mass counterterm
from \eq\nr{deltaM_QCD}, all singular propagators, 
$1/\Delta^{ }_{P_i}$ and $1/\Delta^{ }_{P_i \pm Q}$, 
drop out. 
To achieve this it is important to make use of the fact the certain
differences are of $\rmO(Q)$ and cancel against would-be poles, notably
\ba
 \frac{1}{\Delta^{ }_{P_1 - Q}}
 \biggl[ 
 \frac{1}{\Delta^{ }_{P_1-Q-R}} - \frac{1}{\Delta^{ }_{P_2 - R}}
 \biggr]
 & = & - \frac{
              \omega^{ }_1 + \omega^{ }_2 - q^{ }_0 - 2 r^{ }_0 }
            { (  \omega^{ }_1 + \omega^{ }_2 - q^{ }_0 )
              \,\Delta^{ }_{P_1-Q-R} \,\Delta^{ }_{P_2 - R} }
 \;, \la{trick1} \\ 
 \frac{1}{\Delta^{ }_{P_2 + Q}}
 \biggl[ 
 \frac{1}{\Delta^{ }_{P_2+Q-R}} - \frac{1}{\Delta^{ }_{P_1 - R}}
 \biggr]
 & = & - \frac{
              \omega^{ }_1 + \omega^{ }_2 + q^{ }_0 - 2 r^{ }_0 }
            { ( \omega^{ }_1 + \omega^{ }_2 + q^{ }_0 )
              \,\Delta^{ }_{P_2+Q-R} \,\Delta^{ }_{P_1 - R} }
 \;, \la{trick2}
\ea
where the right-hand sides are non-singular. 
After the elimination of the singular propagators, the non-singular ones
($1/\Delta^{ }_{P_i - R}$, $1/\Delta^{ }_{P_i \pm Q - R}$) can 
be Taylor-expanded in $Q$, with the leading terms 
given by $1/\Delta^{ }_{P-R}$.
The gluonic propagator $1/(Q-R)^2$ can likewise be expanded. 
Left over are tensor integrals of the type
\be
 \int_R \frac{R^{ }_\mu R^{ }_\nu}{(R^2)^{i_1}_{ }\Delta_{P-R}^{i_2} }
 \; = \; 
 A^{ }_{i_1 i_2} \delta^{ }_{\mu\nu} + 
 B^{ }_{i_1 i_2} P^{ }_\mu P^{ }_\nu 
 \;, \quad
 \int_R \frac{R^{ }_\mu}{(R^2)^{i_1}_{ }\Delta_{P-R}^{i_2} }
 \; = \; 
 C^{ }_{i_1 i_2} P^{ }_\mu
 \;, 
\ee
contracted with four-vectors like $\BG(Q)$, $V$ or $Q$, or with 
$\delta^{ }_{\mu i}$, where the index $i$ originates from the operator. 
The tensor integrals can be reduced to scalar ones with the usual 
Passarino-Veltman reduction, e.g.\ 
\ba 
 A^{ }_{i_1 i_2} & = & 
 \frac{c^{ }_{i_1-1,i_2}}{D-1}
 + 
 \frac{
      c^{ }_{i_1-2,i_2} - 2 c^{ }_{i_1-1,i_2-1} + c^{ }_{i_1,i_2 - 2}
      }{4(D-1)M^2} 
 \;, \\
%%%%%
 B^{ }_{i_1 i_2} & = & 
 \frac{c^{ }_{i_1-1,i_2}}{(D-1)M^2}
 + 
 \frac{
      D ( 
      c^{ }_{i_1-2,i_2} - 2 c^{ }_{i_1-1,i_2-1} + c^{ }_{i_1,i_2 - 2}
        )
      }{4(D-1)M^4} 
 \;, \\
 C^{ }_{i_1} & = & 
 \frac{ c^{ }_{i_1,i_2 - 1} - c^{ }_{i_1-1,i_2}  }{2 M^2}
 \;, 
\ea
where 
\be
 c^{ }_{i_1, i_2} \; \equiv \; 
 \int_R \frac{ 1 }{(R^2)^{i_1}_{ }\Delta_{P-R}^{i_2} }
 \;. \la{def_ci1i2}
\ee
Negative powers of $i^{ }_1$ can be dealt with by completing squares, 
e.g.\ 
$
 c^{ }_{-1,i_2} = -2 M^2 c^{ }_{0,i_2} + c^{ }_{0,i_2-1}
$.
After this reduction, we are faced with the integrals  
$
 c^{ }_{0,1}
$, 
$
 c^{ }_{0,2}
$, 
$
 c^{ }_{0,3}
$,
$
 c^{ }_{1,1}
$
and 
$
 c^{ }_{1,2}
$.
In dimensional regularization, these are related by 
\be
 c^{ }_{0,1} = 
   \frac{M^2 c^{ }_{0,2}}{ 1 - \frac{D}{2} }
 \;, \quad
 c^{ }_{0,3} = 
   \frac{ \bigl( 1 - \frac{D}{4} \bigr) c^{ }_{0,2}}{M^2}
 \;, \quad
 c^{ }_{1,1} = 
 \frac{c^{ }_{0,2}}{D-3}
 \;, \quad
 c^{ }_{1,2} = 
 - \frac{c^{ }_{0,2}}{2 M^2}
 \;, \la{vac_master_relation}
\ee
where
$
 c^{ }_{0,2} = 
 \Gamma(2-\frac{D}{2}) 
 / (4\pi)^{\frac{D}{2}}
 / (M^2)^{2 - \frac{D}{2}}_{ }
$.

After inserting the relations 
between the masters integrals, we find that all gauge 
dependence cancels (i.e.\ terms proportional to $1/\xi$, $\xi$, $\xi^2$).
Moreover all terms proportional to $\CA$ cancel
in $D$ dimensions. Terms proportional
to $\CF$ do not cancel, but they come in the same combination
of electric and magnetic fields as the LO result 
in \eq\nr{A_i_QCD_vac_lo}. Furthermore the relative correction,
\be
 M \mathcal{A}_i^\rmii{QCD,vac}  
 \; = \;    
 - \, 2 i \gB^{ } 
 \bigl[ 
   \EA - \EB + \BA % - \BB
 \bigr]
 \, 
 \biggl\{ 
   1 - \gB^2 \CF \int_R \frac{1}{\Delta^2_{P-R}} 
       \frac{D-5}{D-3} 
 \biggr\} 
 + \rmO(\gB^5)
 \;, \la{A_i_QCD_vac_nlo}
\ee
exactly matches that obtained from \eq\nr{N_0_QCD_vac_nlo} after
inserting the relations between the masters
from \eq\nr{vac_master_relation}.
Therefore the ratio defined in \eq\nr{ratio_def1} receives no 
correction at NLO, 
\be
 \frac{ M \mathcal{A}_i^\rmii{QCD,vac} }
      { \mathcal{N}_0^\rmii{QCD,vac}  }
  =    
 -i \gB^{ }
 \bigl[ 
   \EA - \EB + \BA % - \BB
 \bigr]
 \; + \; \rmO(\gB^5)
 \;. \la{a_i_QCD_vac_nlo}
\ee

%%%%%%%%%%%%%%%%%%%%%%%%%%% SECTION %%%%%%%%%%%%%%%%%%%%%%%%%%%%%%%%%%%%%%
%
\section{QCD thermal contribution}
\la{se:qcd_thermal}

The next step is to repeat the computation of \se\ref{se:qcd_vacuum}
at finite temperature. Much remains unchanged, notably the diagrams,
the Dirac contractions, and algebraic steps such 
as completions of squares. What changes is that the gluon four-momentum
is now thermal, $R = (r^{ }_n,\vec{r})$, where $r^{ }_n$ is a bosonic
Matsubara frequency. Integration over $R$ gets replaced by a Matsubara
sum-integral, denoted by $\Tinti{R}$.
As a consequence of the different measure, 
``scaleless'' sum-integrals
no longer vanish, as the temperature sets a new scale. 
In addition, the
symmetry group that permits to eliminate numerators from
sum-integrals
is smaller. 
Most of this 
section concerns how to evaluate these new master sum-integrals.

The first important issue, however, is to note that care is needed 
when Taylor expanding with respect to the external gluon 
four-momentum, which at finite temperature takes the form 
$Q = (q^{ }_n,\vec{q})$, where $q^{ }_n$ is a bosonic
Matsubara frequency. It is well-known, for instance
from the context of Hard Thermal Loop 
effective theories~\cite{htl1,htl2,htl3,htl4}, that after
carrying out the Matsubara sum over $r^{ }_n$, gluon loops
$\sim 1 / [ R^2 (R-Q)^2 ] $ turn into 
structures like 
$
 \sim \nB^{ }(r) /
 \{ r [i q^{ }_n r \pm \vec{q}\cdot\vec{r} + \rmO(Q^2) ] 
 \} 
$. 
We could carry out an analytic continuation to Minkowskian 
frequencies, $i q^{ }_n \to q^{ }_0$. It is then clear that
the result is non-analytic, e.g.\ with a branch
cut in the domain $q > |q^{ }_0|$, leading physically
to the phenomenon of Landau damping. 
Even though the same non-analyticities arise
on the IR side of matching, it is extremely tedious
to track them in an already complicated computation. 
These problems are absent
from the Matsubara zero mode sector, $q^{ }_n = 0$. 
In the language of the Euclidean formulation, 
non-zero Matsubara modes $\sim 2\pi n T$ carry large energies, and therefore
cannot be expanded in; the low-energy mode $q^{ }_n = 0$ suffers
from no such problem. All in all, we therefore restrict to  
the Matsubara zero mode of the external gauge field 
in the thermal computations, {\it viz.} 
\be 
 q^{ }_n = 0
 \;. \la{qn}
\ee 
We note from \eq\nr{fields} that, 
consequently, the electric field denoted by 
$
 \EB 
$
is not available, but this represents no problem,  
because the counterpart 
$
 \EA 
$
remains present.
To avoid confusion, let us stress again that the 
four-momentum of the {\em dynamical} (non-external) 
gauge field, denoted by $R$, does carry all its Matsubara frequencies. 

After this elaboration, 
let us turn to the sum-integrals present, obtained 
after carrying
out the Taylor expansion in $\vec{q}$ and $\vec{v}$.
There are three classes of them:
those sensitive only to the gluon four-momentum ($R$); 
those sensitive only to the heavy quark four-momentum ($P-R$);
and those containing both types of propagators. 
We discuss these in turn. 

The structures only containing the gluon propagator, $\sim 1/R^2$,
vanish in vacuum as scaleless integrals, but are non-zero at finite
temperature. Any spatial momenta appearing in the numerator can be 
eliminated by Passarino-Veltman type reduction but applied in 
$d = 3 - 2\epsilon$ dimensions. Dimensional regularization permits
also to relate a number of integrals, such as 
$
 \int_\vec{r} \frac{\vec{r}^2}{ (R^2)^{i_1} } = 
 \frac{d}{2(i_1-1)} \int_\vec{r}\frac{1}{(R^2)^{i_1 - 1}} 
$.
The remaining 1-loop sum-integrals can be solved in terms
of the Riemann $\zeta$-function, and expansions in $\epsilon$
yield familiar expressions, e.g.\ 
\ba
 \Tint{R}\frac{1}{R^2} & = & 
 \frac{ 2T\,\Gamma( 1 - \frac{d}{2} ) }{ (4\pi)^{d/2} }
 \frac{ \zeta(2-d) }{ (2\pi T)^{2-d} }
 \; = \; \frac{T^2}{12} + \rmO(\epsilon) 
 \;, \la{master_sum0} \\ 
%%%
 \Tint{R} \frac{1}{R^4} 
  & = &  
 \frac{ 2T\,\Gamma( 2 - \frac{d}{2} ) }{ (4\pi)^{d/2} }
 \frac{ \zeta(4-d) }{ (2\pi T)^{4-d} }
 \; = \; 
 \frac{\mu^{-2\epsilon}}{(4\pi)^2}
 \biggl[
   \frac{1}{\epsilon} 
   + 2 \ln\biggl( \frac{\bmu e^{\gammaE}}{4\pi T} \biggr)
   + \rmO(\epsilon)
 \biggr] 
 \;, \la{master_sum1} \\
%%%
 \Tint{R} \frac{1}{R^2 {r}^2}
 & = & 
 \int_\vec{r} \frac{\frac{1}{2} + \nB^{ }(r)}{r^3} 
 \nn 
 & = & - \,
 \frac{ 2T\,\Gamma( 1 - \frac{d}{2} ) }{ (4\pi)^{d/2} }
 \frac{ \zeta(4-d) }{ (2\pi T)^{4-d} }
 \; = \; 
 \frac{2 \mu^{-2\epsilon}}{(4\pi)^2}
 \biggl[
   \frac{1}{\epsilon} 
   + 2 \ln\biggl( \frac{\bmu e^{\gammaE}}{4\pi T} \biggr)
   + 2 + \rmO(\epsilon)
 \biggr] 
 \;. \la{master_sum2}
\ea
The sum-integral in 
\eq\nr{master_sum2} originates in connection 
with mixed structures (see below), and we have shown a
representation in terms of the Bose distribution $\nB^{ }$ 
for later convenience.  

The second class contains only heavy quark propagators, 
\be
 \Tint{R} \frac{R^{ }_\mu R^{ }_\nu ...}{\Delta^{i_1}_{P - R}}
 \;. \la{master_sum3}
\ee
After substituting $R\to P-R$, we are faced with a fermionic
Matsubara sum. Carrying it out, the thermal part of the result 
comes with the Fermi distribution~$\nF^{ }$, 
which is exponentially
suppressed by $\sim e^{-M/T}$. Therefore, \eq\nr{master_sum3} can 
be replaced by its vacuum part, 
$
 \int_R {R^{ }_\mu R^{ }_\nu ...}/{\Delta^{i_1}_{P - R}}
$, 
and it then evaluates to the same value as in \se\ref{se:qcd_vacuum}.

The third class contains mixed structures. 
To see what happens with them, we note that 
\ba
 \Tint{R} \frac{ \phi(\vec{r}) }{(R^2+\lambda^2)\Delta^{ }_{P-R} }
 & = &  
 \int_R \frac{ \phi(\vec{r}) }{(R^2+\lambda^2)\Delta^{ }_{P-R} }
 \nn 
%%%%%%%%%%
 & + & 
 \int_{\vec{r}}
 \frac{\nB^{ }(\epsilon^{ }_r)}{2 \epsilon^{ }_r }
 \biggl[
   \frac{ \phi(\vec{r}) }{\epsilon^2_{pr} 
   - (i \omega^{ }_n + \epsilon^{ }_r)^2} 
  + 
   \frac{ \phi(\vec{r}) }{\epsilon^2_{pr}
   - (i \omega^{ }_n - \epsilon^{ }_r)^2} 
 \biggr]^{ }_{ }
 \la{master_sum4} \\
%%%%%%%%%%
 & - & 
 \int_{\vec{r}}
 \frac{\nF^{ }(\epsilon^{ }_{pr})}{2 \epsilon^{ }_{pr}}
 \biggl[
   \frac{ \phi(\vec{r}) }{\epsilon^2_{r}
   - (i \omega^{ }_n + \epsilon^{ }_{pr})^2} 
  + 
   \frac{ \phi(\vec{r}) }{\epsilon^2_{r}
   - (i \omega^{ }_n - \epsilon^{ }_{pr})^2} 
 \biggr]^{ }_{ }
 \;, \nonumber
\ea
where we denoted $P = (\omega^{ }_n,\vec{p})$, 
$ 
 \epsilon^{ }_r \equiv \sqrt{r^2 + \lambda^2}
$
and
$ 
 \epsilon^{ }_{pr} \equiv \sqrt{(\vec{p}-\vec{r})^2 + M^2}
$.
Taking derivatives with respect to $\lambda^2$ and $M^2$ 
permits to generate powers of propagators. 
The first term on the right-hand side 
of \eq\nr{master_sum4} is a vacuum integral, and reproduces the 
effects found in \se\ref{se:qcd_vacuum}. 
The last term is 
exponentially suppressed like the thermal effects originating 
from \eq\nr{master_sum3}, and can be omitted. 
Relevant contributions  
originate from the middle term of \eq\nr{master_sum4}. 
The same exercise can be repeated for the case that $r^{ }_n$ appears
in the numerator, and then the middle term reads
\be
 \Tint{R} \frac{ \phi(\vec{r})\, r^{ }_n }{(R^2+\lambda^2)\Delta^{ }_{P-R} }
 \; \supset \;  
 \int_{\vec{r}}
 \frac{\nB^{ }(\epsilon^{ }_r) }{2 \epsilon^{ }_r }
 \biggl[
   \frac{ \phi(\vec{r})\, i \epsilon^{ }_r 
   }{\epsilon^2_{pr}
   - (i \omega^{ }_n + \epsilon^{ }_r)^2} 
  - 
   \frac{ \phi(\vec{r})\, i \epsilon^{ }_r
  }{\epsilon^2_{pr}
   - (i \omega^{ }_n - \epsilon^{ }_r)^2} 
 \biggr]^{ }_{ }
 \;.
 \la{master_sum5}
\ee

Subsequently, we set the heavy quarks on-shell, 
$\omega^{ }_n \to i \epsilon^{ }_p$ 
like in \eq\nr{on-shell},\footnote{%
 The precise justification for this in the thermal context is 
 provided in \se\ref{se:nonrel}.
 } 
and expand the result in 
$\vec{v} = \vec{p} / \epsilon^{ }_p$ and $T/M$, 
where the temperature originates from the fact that $\epsilon^{ }_r\sim T$, 
as dictated by the Bose distribution. 
In this way we find that, effectively, 
\ba
 \frac{c^{ }_0 + c^{ }_1\, r^{ }_n}{\Delta^{ }_{P-R}} 
 & \stackrel{\rmii{mixed term}}{\longrightarrow} & 
 \frac{ c^{ }_0\, \vec{r}\cdot\vec{v}
   + c^{ }_1 \, i \epsilon_r^2 
     }{M \epsilon_r^2}
      + ...
 \;, \la{expa1} \\ 
%%%%
 \frac{c^{ }_0 + c^{ }_1\,r^{ }_n}{\Delta^{2}_{P-R}} 
 & \stackrel{\rmii{mixed term}}{\longrightarrow} & 
 \frac{ c^{ }_0 
 + c^{ }_1\, 2 i \vec{r}\cdot\vec{v} 
     }{2 M^2 \epsilon_r^2}
      + ...
 \;, \la{expa2} \\ 
%%%%
 \frac{c^{ }_0 + c^{ }_1\,r^{ }_n}{\Delta^{3}_{P-R}} 
 & \stackrel{\rmii{mixed term}}{\longrightarrow} & 
 \frac{ c^{ }_0\, 3  \vec{r}\cdot\vec{v}
     + c^{ }_1\, i  \epsilon_r^2 
     }{4 M^3 \epsilon_r^4}
      + ...
 \;, \la{expa3}
\ea
appearing together with 
$ 
 \int_{\vec{r}}
 {\nB^{ }(\epsilon^{ }_r) } / {(2 \epsilon^{ }_r )}
$
that was factored out in \eqs\nr{master_sum4} and \nr{master_sum5}. 

A few further remarks are in order. First, 
we note that if $r_n^2$ appears in the numerator, 
it can be written as 
$
 r_n^2 = R^2 + \lambda^2 - \epsilon_r^2
$, 
and thus represented as a linear combination of 
the structures that were already considered. A case 
to watch out for is if the function $\phi(\vec{r})$, 
perhaps in combination with the right-hand sides of 
\eqs\nr{expa1}--\nr{expa3}, leads to a spatial momentum squared, 
e.g.\ $r^{ }_i r^{ }_j \to \delta^{ }_{ij} {r}^2 / d$. We may now write
$
 r^2 = \epsilon_r^2 - \lambda^2
$. 
If this appears in a structure with a quadratic gluon propagator, 
$
 1/R^4 = - \lim_{\lambda\to 0} {{\rm d}} / {{\rm d}\lambda^2}
 \{ 1/(R^2 + \lambda^2) \}
$, 
then the derivative can act on the numerator as well, implying that 
$r^2 /[R^4 (\epsilon_r^2)^{ i_1}] \to 
 1 / [R^2 (\epsilon_r^2)^{ i_1}]
 +
 1/[R^4 (\epsilon_r^2)^{ i_1 - 1} ]
$.
% A ``rule of thumb'' for this behaviour is to think of 
% $1/\epsilon_r^2$ as equivalent to $1/(i r^{ }_n)^2$,\footnote{% 
%  More precisely, we mean $1/\Lambda^2_{P-R}$, 
%  where $\Lambda$ is the static inverse 
%  fermionic propagator from \eq\nr{Lambda}.
%  }
% and then to write $r^2 = R^2 + (i r^{ }_n)^2$.

To summarize, when we send $\lambda\to 0$, thermal parts of 
mixed sum-integrals can be represented in terms of 
\eqs\nr{master_sum0}--\nr{master_sum2}. 
After inserting all this to the diagrams
of \fig\ref{fig:QCD}, we obtain results for the contribution
from thermal scales. We postpone their discussion till the end
of \se\ref{se:nonrel}, where the 
main result, given in \eq\nr{A_i_nr_nlo_expl}, 
is obtained in a different way. 

%%%%%%%%%%%%%%%%%%%%%%%%%%% SECTION %%%%%%%%%%%%%%%%%%%%%%%%%%%%%%%%%%%%%%
%
\section{Non-relativistic determination of the thermal contribution}
\la{se:nonrel}

The purpose of this section is to re-derive the result 
of \se\ref{se:qcd_thermal} in a different way. 
For practical applications, there is thus nothing new; 
however, on the formal side, 
we hope that an independent derivation
can serve as a crosscheck and an illustration of the general methodology. 
Moreover this approach brings us in several ways
rather close to \se\ref{se:ir}.

The idea is to use a non-relativistic effective theory
for the computation. 
Whereas full QCD has two scales that we treated separately, 
$M$ in \se\ref{se:qcd_vacuum} and $T$ in \se\ref{se:qcd_thermal}, 
the scale $M$ has essentially been eliminated from the 
effective theory. 
This permits to simplify some aspects of the computation 
(for instance, spin plays a trivial role and
Dirac matrices do not appear), even if there is also an overhead, 
namely an increased number of elementary vertices. 

The Euclidean action of the non-relativistic theory reads
\be
 S^{ }_\rmii{E} = 
 \int_X \theta^\dagger_{ }
 \, \biggl( 
      D^{ }_0 + M - 
     \frac{\vec{D}^2 + 
    c^{ }_\B\, \gB^{ } \vec{\sigma}\cdot\vec{B}}{2 M} + \ldots 
    \biggr) \, 
 \theta
 \;, \la{S_nr}
\ee
where $\int_X \equiv \int\!{\rm d}\tau\! \int_\vec{x}$,  
and $c^{ }_\B = 1 + \rmO(\gB^2)$
is a matching coefficient. Spin-dependent effects are
mass-suppressed and do not contribute to our actual computation, 
however we have shown the term multiplied by $c^{ }_\B$
because it is needed in 
\se\ref{se:discussion}. 
Even if we mentioned
above that the scale $M$ has
essentially been eliminated, it is important for
thermal computations to keep the rest mass explicit in \eq\nr{S_nr}, 
as otherwise Boltzmann factors $e^{-M/T}$ go amiss. 

The reason for an increased number of vertices is that 
\eq\nr{S_nr} contains not only a linear appearance of gauge fields, 
as is the case 
in the heavy-quark part of the QCD action, but higher powers as well. 
% Indeed, apart from the single-gluon structure
% \be
%  S^{ }_\rmii{E} \supset
%  - i g 
%  \Tint{P^{ }_1,P^{ }_2,Q} \hspace*{-7mm} 
%  \deltabar(Q + P^{ }_2 - P^{ }_1)\, 
%  \theta^\dagger(P^{ }_1) \tilde\gamma^{ }_{\mu}(P^{ }_1 + P^{ }_2)
%   A^{ }_\mu(Q)\, \theta(P^{ }_2)
%  \;, \quad
%  \tilde\gamma^{ }_{\mu}(P) 
%  \; \equiv \; 
%  \delta^{ }_{\mu 0} 
%  - \frac{i \delta^{ }_{\mu j} p^{ }_{j} }{2M} + ... 
%  \;, 
% \ee
% the term suppressed by $1/M$ 
% in \eq\nr{S_nr}
% leads to a two-gluon-vertex, 
% and this turns out to contribute to our computation. 
Likewise, the spatial Noether current, 
\be
 J^\rmii{HQET}_i \; = \; - \,  
 \frac{ \theta^\dagger 
 ( i \overleftrightarrow{D}^{ }_{\!\!i}\,) \theta }{ 2M } 
 + \rmO\biggl( \frac{1}{M^2} \biggr)
 \;, 
\ee
involves terms with and without gauge fields. We note that all terms
of $\rmO(1/M)$ and $\rmO(\vec{v}/M)$ need to be included, as the acceleration
is multiplied by $M$ in \eq\nr{matching}. 

In the non-relativistic theory, free propagators take the form
\be
 \langle \theta(P^{ }_1) \, \theta^\dagger(P^{ }_2) \rangle
 = 
 \frac{\deltabar(P^{ }_1 - P^{ }_2)}
 {\Omega^{ }_{P^{ }_1}}
 \;, \quad
 \Omega^{ }_{P^{ }_1}
 \; \equiv \; 
 i \omega^{ }_{1n} + \epsilon^{ }_{p^{ }_1} 
 \;, 
\ee
where $P^{ }_{1} = (\omega^{ }_{1n},\vec{p}^{ }_1)$, 
$\omega^{ }_{1n}$ denotes a fermionic Matsubara frequency, 
$ \epsilon^{ }_{p_1} =  
  M
  + {\vec{p}_1^2}/({2M}) + ... $, 
and $\Tinti{P^{ }_1}\deltabar(P^{ }_1) = 1$. 
We assume all dependence on $1/M$ to be 
Taylor-expanded to a given order.
In the end,  
propagators therefore appear in a static form,  
i.e.\ as inverses of 
\be
 \Lambda^{ }_{P^{ }_1}
 \; \equiv \; 
 i \omega^{ }_{1n} + M 
 \;. \la{Lambda}
\ee  

Let us start with LO computations. 
For the denominator, where the operator reads 
$
 \int^{ }_\vec{x}
 \theta^\dagger(0,\vec{x})\theta(0,\vec{x})
 = 
 \Tinti{P^{ }_3,P^{ }_4} 
 \theta^\dagger(P^{ }_3)\theta(P^{ }_4) 
 \delta^{(3)}_{ }(\vec{p}^{ }_3 - \vec{p}^{ }_4)
$, 
\eq\nr{partition} leads to 
\be
 T\sum_{\omega^{ }_{1n}} e^{\frac{i\beta\omega_{1n}}{2}}
 \, 
 T\sum_{\omega^{ }_{2n}} e^{\frac{i\beta\omega_{2n}}{2}}
 \frac{2\,
      \delta^{(3)}(\vec{p}^{ }_1 - \vec{p}^{ }_2)
      }{(i \omega^{ }_{1n} + M )(i \omega^{ }_{2n} + M)}
 + \rmO\biggl(\frac{1}{M}\biggr)
 \;. \la{partition_nr_lo}
\ee
The Matsubara sums yield $e^{-M/T}$. Factoring out this exponential, 
as well as $ \delta^{(3)}(\vec{p}^{ }_1 - \vec{p}^{ }_2) $, 
the ``amplitude''
corresponding to \eq\nr{denominator_def1} is now extracted as 
\be
 \mathcal{N}^\rmii{HQET}_{0} 
 = 2  + \rmO(\gB^2)
 \;. \la{N_0_HQET_lo}
\ee

%%%%%%%%%%%%%%%%%%%%%%%%% FIGURE %%%%%%%%%%%%%%%%%%%%%%%%%%%%%%%%%%%%%%%%%
%
\begin{figure}[t]
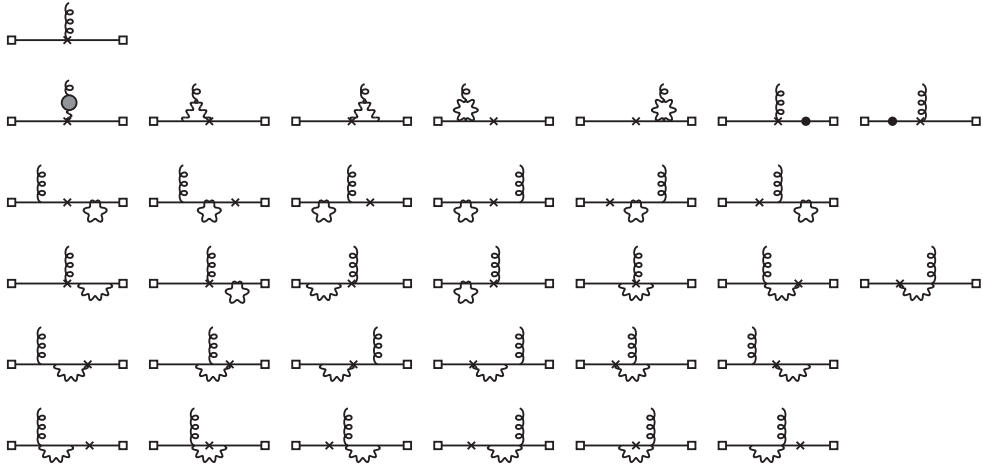


\hspace*{0.5cm}%
\begin{minipage}[c]{15cm}
\begin{eqnarray*}
 &&
 \hspace*{-1.0cm}
%%%%%%%%%%
 \procNRa
%%%%%%%%%%
 \nn[2mm] 
 &&
 \hspace*{-1.0cm}
 \procNRc
 \hspace*{0.20cm}
 \procNRe
 \hspace*{0.20cm}
 \procNRg
 \hspace*{0.20cm}
 \procNRee
 \hspace*{0.20cm}
 \procNRgg
 \hspace*{0.20cm}
%%%%%%%%%%
% \nn[2mm] 
% &&
% \hspace*{-1.0cm}
 \procNRaX
 \hspace*{0.20cm}
 \procNRbX
%%%%%%%%%%
 \nn[2mm] 
 &&
 \hspace*{-1.0cm}
 \procNRaA
 \hspace*{0.20cm}
 \procNRaD
 \hspace*{0.20cm}
 \procNRaF
 \hspace*{0.20cm}
 \procNRbA
 \hspace*{0.20cm}
 \procNRbD
 \hspace*{0.20cm}
 \procNRbF
%%%%%%%%%%
 \nn[2mm] 
 &&
 \hspace*{-1.0cm}
 \procNRaAA
 \hspace*{0.20cm}
 \procNRaAB
 \hspace*{0.20cm}
 \procNRaEE
 \hspace*{0.20cm}
 \procNRaEF
 \hspace*{0.20cm}
 \procNRaBB
 \hspace*{0.20cm}
%%%%%%%%%%
% \nn[2mm] 
% &&
% \hspace*{-1.0cm}
 \procNRaAS
 \hspace*{0.20cm}
 \procNRaAT
%%%%%%%%%%
 \nn[2mm] 
 &&
 \hspace*{-1.0cm}
 \procNRaAM
 \hspace*{0.20cm}
 \procNRaAN
 \hspace*{0.20cm}
 \procNRaAO
 \hspace*{0.20cm}
 \procNRaAP
 \hspace*{0.20cm}
 \procNRaAQ
 \hspace*{0.20cm}
 \procNRaAR
%%%%%%%%%%
 \nn[2mm] 
 &&
 \hspace*{-1.0cm}
 \procNRaAU
 \hspace*{0.20cm}
 \procNRaAV
 \hspace*{0.20cm}
 \procNRaAW
 \hspace*{0.20cm}
 \procNRaAX
 \hspace*{0.20cm}
 \procNRaAY
 \hspace*{0.20cm}
 \procNRaAZ
% \nn[-10mm]
\end{eqnarray*}
\end{minipage}

\caption[a]{\small 
 The {\em additional}
 LO and NLO graphs contributing to the 3-point correlator 
 in the non-relativistic description of \se\ref{se:nonrel}. 
 The notation is the same as in \fig\ref{fig:QCD}. 
 We note that in a thermal medium, gluon tadpoles give a finite
 contribution, proportional to $T^2$, and must thus be included.} 
 \la{fig:NR}
\end{figure}
%
%%%%%%%%%%%%%%%%%%%%%%%%%%%%%%%%%%%%%%%%%%%%%%%%%%%%%%%%%%%%%%%%%%%%%%%%%%

For the numerator, the momentum space operator becomes
\ba
% & & \hspace*{-1.0cm} 
 \partial^{ }_0 \int_\vec{x} J^\rmii{HQET}_i 
 & = &  
 \Tint{P^{ }_3,P^{ }_4} \hspace*{-5mm} 
 i (\omega^{ }_{4n} - \omega^{ }_{3n})\,
 \theta^\dagger(P^{ }_3)
 \, \frac{p^{ }_{3i} + p^{ }_{4i}}{2M}\, \theta(P^{ }_4)\,
 \delta^{(3)}_{ }(\vec{p}^{ }_4 - \vec{p}^{ }_3) 
 \\
%%%%
 & - &  
 \Tint{Q, P^{ }_3,P^{ }_4} \hspace*{-8mm}
 i (q^{ }_n + \omega^{ }_{4n} - \omega^{ }_{3n})\,
 \theta^\dagger(P^{ }_3)
 \, \frac{g A^{ }_{i}(Q)}{M}\, \theta(P^{ }_4)\,
 \delta^{(3)}_{ }(\vec{q} + \vec{p}^{ }_4 - \vec{p}^{ }_3)
 + ... 
 \;, \hspace*{6mm}  \nonumber
%%%%
\ea
where corrections start at $\rmO(1/M^2)$. 
We get a contribution from three diagrams
at leading order, 
illustrated on the first rows of \figs\ref{fig:QCD} and~\ref{fig:NR}.
There is an issue with singularities, 
similar to that discussed around \eq\nr{cancel_1}, 
but with non-relativistic propagators 
the cancellation is simpler,\footnote{% 
  To justify the use of on-shell conditions here, 
  i.e.\ $\omega^{ }_{in} = i \epsilon^{ }_{p_i}$, 
  we note that by adding and subtracting a term,
  e.g.\ 
  $\omega^{ }_{1n}/(i \omega^{ }_{1n} + \epsilon^{ }_{p^{ }_1} )
  = -i 
  + i \epsilon^{ }_{p^{ }_1} /(i \omega^{ }_{1n} + \epsilon^{ }_{p^{ }_1} )$
  in the term containing $1/\Omega^{ }_{P^{ }_2 + Q}$, 
  we are left 
  with a Matsubara sum like in 
  \eq\nr{partition_nr_lo}, but with one of the terms
  (here, $-i$) 
  being independent of one of the summation variables 
  (here, $\omega^{ }_{1n}$).
  These terms vanish 
  in connection with the exponentials.
 } 
\be
 \frac{q^{ }_n + \omega^{ }_{2n} - \omega^{ }_{1n} }
 {\Omega^{ }_{P^{ }_1 - Q}}
 = i 
 \;, \quad 
 \frac{q^{ }_n + \omega^{ }_{2n} - \omega^{ }_{1n} }
 {\Omega^{ }_{P^{ }_2 + Q}}
 = - i 
 \;. \la{ids_nr}
\ee

After inserting the small-momentum approximations from \eq\nr{taylor}, 
setting $q^{ }_n = 0$ for the external gauge field 
as explained around \eq\nr{qn}, 
Taylor-expanding, factoring out external states like 
around \eq\nr{N_0_HQET_lo}, 
and making use of the notation in \eq\nr{fields}, 
we find
\be
 M\, \mathcal{A}_i^\rmii{HQET}
 \; = \; 
 2 \gB^{ } 
 \bigl[ 
   \EA % - \EB
 + \BA % - \BB
 \bigr] 
 + \rmO(\gB^3)
 \;. \la{A_i_HQET_lo}
\ee
The ratio of \eqs\nr{A_i_HQET_lo} and \nr{N_0_HQET_lo}
yields a structure similar to 
the Lorentz force in \eq\nr{lorentz_em}.

Proceeding to NLO, we start with the denominator, 
deferring the discussion of technical details to the 
numerator. 
% The additional topologies, supplementing those in 
% the full QCD computation, can be obtained from the diagrams
% in \fig\ref{fig:NR}, by eliminating the external gluon line. 
Evaluating the NLO correction at the tree-level on-shell point, 
the final result reads 
\be
\frac{ \mathcal{N}^\rmii{HQET}_0  }
     { 2 } 
 =  1 - \gB^2 \CF \Tint{R} \biggl[ 
      \frac{1}{R^2\Lambda^2_{P-R}}
      + 
      \frac{1 - \xi}{R^4} 
 \biggr]
 + \rmO(\gB^4) 
 \;, \la{N_0_HQET_nlo}
\ee
where $\Lambda$ is the inverse static propagator from \eq\nr{Lambda},
and $\xi$ is a gauge parameter. 
After the insertion of master sum-integrals from 
\eqs\nr{master_sum1}, \nr{master_sum2} and \nr{master_sum6}, 
we obtain
\be
\frac{ \mathcal{N}^\rmii{HQET}_0  }
     { 2 } 
 = 
 1 
 - 
  \frac{ \gB^2 \CF \,\mu^{-2\epsilon}}{(4\pi)^2}
  \biggl\{ 
   \bigl( 3  - \xi  \bigr) 
   \biggl[ \frac{1}{\epsilon} + 
      2 \ln\biggl( \frac{\bmu e^{\gammaE}}{4\pi T} \biggr)  \biggr] 
      + 4
  \biggr\}
  + \rmO(\epsilon\,\gB^2,\gB^4)   
 \;. \la{N_0_HQET_nlo_expl}
\ee
 The gauge parameter appears because \eq\nr{partition} is not explicitly gauge 
 invariant, and its ultimate cancellation serves as an important 
 crosscheck of the computation. 

Turning to the numerator, 
let us first discuss the mass counterterm.
In order to cancel all singular propagators
($1/ \Omega^{ }_{P_i}$, $1/ \Omega^{ }_{P_i \pm Q}$), the mass counterterm
needs to be chosen such that we are in an on-shell scheme. 
In the non-relativistic theory, the counterterm is analogous to 
that in \eq\nr{deltaM_QCD} but now with a thermal sum-integral
($M_\rmii{B}^{ } = M + \delta M$), 
\be 
 \delta M = - \gB^2\CF \Tint{R}
 \biggl[
    \frac{1}{ R^2 \Omega^{ }_{P-R}}
  +
    \frac{D-1}{2 M R^2} 
 \biggr]
 \; + \; \rmO(\gB^4) 
 \;. \la{deltaM_NR}
\ee
In the main computation it is convenient to use 
this in unexpanded form, in order to guarantee that 
the cancellation outlined in  
\eqs\nr{simpl_nr_1}, 
\nr{simpl_nr_2} 
takes place at an early stage, but we note
in passing that if we wanted an explicit value, we could expand the 
propagator as
\ba
 \Tint{R} \frac{1}{R^2\Omega^{ }_{P-R}} & = & 
 \Tint{R} \frac{1}{R^2} 
 \biggl( \frac{1}{\Lambda^{ }_{P-R}}
 - \frac{\vec{r}^2}{2M} \frac{1}{\Lambda^2_{P-R}}
 \biggr) 
 + \rmO\biggl(\vec{v}^2,\frac{T^3}{M^2}\biggr)
 \;. \hspace*{6mm}
\ea
The sums can be performed (cf.\ \eqs\nr{sum1}, \nr{master_sum6}) and 
subsequently related to that in \eq\nr{master_sum0}. The upshot is  
that the vacuum pole mass is shifted by a well-known thermal
correction~\cite{dhr}, 
\be
 M |^{ }_\T = M |^{ }_\rmii{$T=0$} +  \frac{g^2 T^2 \CF}{12 M}
 \;. 
\ee 

With the mass counterterm from \eq\nr{deltaM_NR}, 
the cancellation of singular propagators requires the use of identities
analogous to \eqs\nr{trick1} and \nr{trick2}.
In the non-relativistic theory, their form is simplified to 
\ba
 \frac{1}{\Omega^{ }_{P_1 - Q}}
 \biggl[ 
 \frac{1}{\Omega^{ }_{P_1-Q-R}} - \frac{1}{\Omega^{ }_{P_2 - R}}
 \biggr]
 & = & - \frac{ 1 }
            { \Omega^{ }_{P_1-Q-R}\, \Omega^{ }_{P_2 - R} }
 \;, \la{simpl_nr_1} \\ 
 \frac{1}{\Omega^{ }_{P_2 + Q}}
 \biggl[ 
 \frac{1}{\Omega^{ }_{P_2+Q-R}} - \frac{1}{\Omega^{ }_{P_1 - R}}
 \biggr]
 & = & - \frac{ 1 } 
            { \Omega^{ }_{P_2+Q-R}\, \Omega^{ }_{P_1 - R} }
 \;. \la{simpl_nr_2}
\ea

After the cancellation of singularities, 
we can Taylor-expand the non-singular propagators, 
obtaining powers of $1/\Lambda^{ }_{P-R}$. 
As explained around \eq\nr{qn}, the Taylor expansion of the 
gluon propagator $1/(Q-R)^2$ is sensible only with respect
to spatial momentum $\vec{q}$, 
so we restrict to the Matsubara zero mode $q^{ }_n = 0$.
In order to handle the large number of diagrams, 
shown in \figs\ref{fig:QCD} and \ref{fig:NR},  
and the many terms
generated by their Taylor expansions, we have made extensive
use of FORM~\cite{form}. 

% To reduce the number of terms, 
% it may be helpful to eliminate terms like
% $
%  r^{ }_n / \Omega^{ }_{P-R}
% $
% or 
% $
%  r^{ }_n / \Lambda^{ }_{P-R}
% $
% by adding and subtracting  

As far as the Matsubara sums go, we need to replace 
\eq\nr{master_sum4} with its 
non-relativistic counterpart. The sum now reads 
\ba
 && \hspace*{-1.5cm}
 T\sum_{r_n} \frac{1}{(R^2 + \lambda^2) \Lambda^{ }_{P-R}}
 \; = \; 
 T\sum_{r_n} \frac{1}{(r_n^2 + \epsilon_r^2) 
                      [ i(\omega^{ }_n - r^{ }_n) + M ]}
 \nn 
 & = &  
 \frac{1}{(i\omega^{ }_n + M )^2 - \epsilon_r^2}
 \biggl\{ 
    \frac{i\omega^{ }_n + M }{\epsilon^{ }_r}
    \biggl[ \frac{1}{2} + \nB^{ }( \epsilon^{ }_r )
    \biggr]
   - 
    \biggl[ \frac{1}{2} - \nF^{ }( M ) 
    \biggr]
 \biggr\} 
 \;. \hspace*{6mm} \la{sum1}
\ea
Taking a derivative with respect to $ M $ and going subsequently 
on-shell, $\omega^{ }_n\to i M $, leads to 
\be
 T\sum_{r_n} \frac{1}{( R^2 + \lambda^2 )\Lambda^{2}_{P-R}} 
 \biggr|^{ }_{\omega^{ }_n = i M }
 = 
 \frac{1}{\epsilon^{2}_r}
 \biggl[
   \frac{\frac{1}{2} + \nB^{ }(\epsilon^{ }_r)}{\epsilon^{ }_r} 
   + \nF'( M )
 \biggr]
 \;. \la{master_sum6}
\ee
For $M \gg T$, the term proportional to $\nF'$ is exponentially 
suppressed, so after $\lambda\to 0$ 
we are left over with the purely bosonic term
in \eq\nr{master_sum2}. 
For non-trivial numerators, 
e.g.\ 
\be
 \Tint{R} \frac{r^{ }_i r^{ }_j}{(R^2)^{i_1}_{ }\Lambda_{P-R}^{i_2} }
 \;, \la{tensors}
\ee
the discussion in the paragraph below \eq\nr{expa3} applies. 

As a final technical remark, 
we note that gluon self-energy
contributions, shown on the second rows of 
\figs\ref{fig:QCD} and \ref{fig:NR}, 
do not need to be included. The reason is that
they yield precisely the same contribution as in the 
IR description, whose graphs are shown in \fig\ref{fig:IR}. 
Therefore the self-energy contribution drops out in the 
matching step, discussed in \se\ref{se:discussion}. 

All in all the thermal NLO result for the numerator can be expressed as 
\ba
 \frac{ M \mathcal{A}_i^\rmii{HQET} }
      {  2   }
 & = & 
 \gB^{ } \, 
 \bigl[ 
   \EA + \BA % - \BB
 \bigr]
 \biggl\{ 
 1 
 - 
  \gB^2 \CF \Tint{R} 
      \biggl[
       \frac{1}{R^2\Lambda^2_{P-R}}
       + 
       \frac{ 1 -  \xi }{R^4} 
      \biggr]
 \biggr\} 
 \nn
%%%%%%
 & + & 
 \frac{\gB^3\CA}{2}
      \Tint{R} \biggl\{ 
       \biggl[ 
        \frac{1}{R^2\Lambda^2_{P-R}}
       + 
       \frac{(d-3)(\xi  - 1) - 2}{R^4} 
       \biggr]
        \EA
%  \nn
%%%%%%
%  & & \hspace*{1.4cm} 
      \, + \, 
      \biggl[
       \frac{1 - \frac{2}{d} }{R^2\Lambda^2_{P-R}}
       -
       \frac{\frac{2}{d}}{R^4} 
      \biggr]
         % \Bigl[
             \BA 
         % - \BB \Bigr]
      \biggr\}
 \nn[2mm] 
%%%%%%%
 & + & 
  \mbox{(gluon self-energy)}
      \; + \; \rmO(\gB^5)
 \;. \la{A_i_nr_nlo}
\ea 
The term proportional to $\CF^{ }$ agrees with 
\eq\nr{N_0_HQET_nlo}, and thus drops out in the ratio
considered in \eq\nr{matching}. 
The coefficient of 
$
  \BA % - \BB ]
$
on the second row of \eq\nr{A_i_nr_nlo}
cancels exactly, 
given that the sum-integrals in 
\eqs\nr{master_sum1} and \nr{master_sum2} differ by 
a factor $d/2 - 1$.
When the same relation is inserted into the coefficient
of 
$
  \EA
$, 
the result does not cancel but is proportional to $d-3$. 
Because the coefficient function has a pole $\sim 1/\epsilon$, 
this leaves over a finite contribution,  
\ba
 \frac{ M \mathcal{A}_i^\rmii{HQET} }
      {  \mathcal{N}_0^\rmii{HQET}}
 & = & 
  \gB^{ } \, \biggl\{ 
   \EA 
  \, \biggl[
  1 +  
 \frac{\gB^2\CA (3 - \xi) }{(4\pi)^2}
     \biggr]
 + \BA % - \BB
 \biggr\} 
 \nn[2mm]
%%%%%%
 & + & 
  \mbox{(gluon self-energy)}
      \; + \; \rmO(\epsilon\,\gB^3,\gB^5)
 \la{A_i_nr_nlo_expl}
 \;.
\ea
Amusingly, a finite term proportional to 
$3 - \xi$ is familiar from rescalings discussed in the context of
the effective potential for $\BG^{ }_0$, 
cf.\ \eqs(3.17-18) of ref.~\cite{cka}.

%%%%%%%%%%%%%%%%%%%%%%%%%%% SECTION %%%%%%%%%%%%%%%%%%%%%%%%%%%%%%%%%%%%%%
%
\section{Infrared side of the matching}
\la{se:ir}

%%%%%%%%%%%%%%%%%%%%%%%%% FIGURE %%%%%%%%%%%%%%%%%%%%%%%%%%%%%%%%%%%%%%%%%
%
\begin{figure}[t]
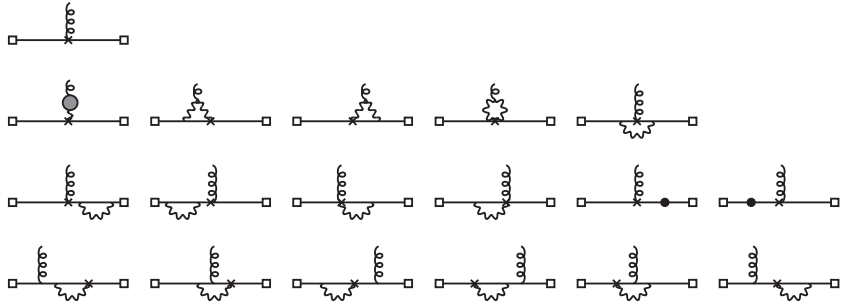


\hspace*{0.5cm}%
\begin{minipage}[c]{15cm}
\begin{eqnarray*}
 &&
 \hspace*{-1.0cm}
%%%%%%%%%%
 \procNRa
%%%%%%%%%%
 \nn[2mm] 
 &&
 \hspace*{-1.0cm}
 \procNRc
 \hspace*{0.20cm}
 \procNRe
 \hspace*{0.20cm}
 \procNRg
 \hspace*{0.20cm}
 \procNReE
 \hspace*{0.20cm}
 \procNRaBB
%%%%%%%%%%
 \nn[2mm] 
 &&
 \hspace*{-1.0cm}
 \procNRaAA
 \hspace*{0.20cm}
 \procNRaEE
 \hspace*{0.20cm}
 \procNRaASS
 \hspace*{0.20cm}
 \procNRaATT
 \hspace*{0.20cm}
 \procNRaX
 \hspace*{0.20cm}
 \procNRbX
%%%%%%%%%%
 \nn[2mm] 
 &&
 \hspace*{-1.0cm}
 \procNRaAM
 \hspace*{0.20cm}
 \procNRaAN
 \hspace*{0.20cm}
 \procNRaAO
 \hspace*{0.20cm}
 \procNRaAP
 \hspace*{0.20cm}
 \procNRaAQ
 \hspace*{0.20cm}
 \procNRaAR
% \nn[-10mm]
\end{eqnarray*}
\end{minipage}

\caption[a]{\small 
 The LO and NLO graphs contributing to the 3-point correlator 
 in the IR description of \se\ref{se:ir}. 
 The notation is the same as in \fig\ref{fig:QCD}. } 
 \la{fig:IR}
\end{figure}
%
%%%%%%%%%%%%%%%%%%%%%%%%%%%%%%%%%%%%%%%%%%%%%%%%%%%%%%%%%%%%%%%%%%%%%%%%%%

 In the preceding sections, we have computed the contributions of 
 the vacuum ($\sim M$) and thermal ($\sim \sqrt{MT},T$) scales to 
 the left-hand side of \eq\nr{matching}. The last ingredient needed for 
 matching is to determine the right-hand side of \eq\nr{matching},
 by making use of the IR description. 
This is defined by restricting to a strictly static HQET action, 
\be
 S^{ }_\rmii{E} \; \equiv \; 
 \int_X \theta^\dagger \, \bigl( D^{ }_0 + M \bigr) \, \theta
 \;. \la{SE_ir}
\ee
Consequently heavy quark propagators are 
straight Wilson lines in the time direction. 
In momentum space, the inverse propagator takes 
the form of \eq\nr{Lambda}.

We note that the IR physics of 
the thermal gluon sector is non-trivial, 
leading to non-analyticities 
as discussed around \eq\nr{qn}
and requiring
resummations in order to generate a consistent weak-coupling series. However,
since we are matching two different computations, these IR issues drop out, 
as long as they have been treated in the same way on both sides of the 
matching. We implement this 
by restricting to $q^{ }_n = 0$
and
by carrying out unresummed computations throughout.
Even after these simplifications, nice crosschecks do remain, 
in particular
that the gauge-dependent electric field normalization visible on 
the second row of \eq\nr{A_i_nr_nlo} 
is reproduced in $d$ dimensions (cf.\ \eq\nr{F_i_IR_nlo}). 

The operator for the denominator reads 
$\int_\vec{x}\theta^\dagger\theta$, and that 
for the numerator $\int_\vec{x} F^\rmii{IR}_i$, 
where $ F^\rmii{IR}_i $ is given in \eq\nr{F_IR}.
As discussed below \eq\nr{F_IR}, in this description $\vec{v}$ appears 
as an external parameter, whose value is fixed later on 
from a separate consideration (cf.\ \se\ref{se:discussion}).
% In momentum space we may write
% the field-strength part of $F^\rmii{IR}_i$ as 
% \be
%  F^{ }_{i \mu}(Q) 
%  = i\, \bigl[\, Q^{ }_i A^{ }_\mu (Q) - Q^{ }_\mu A^{ }_i (Q) \,\bigr]
%  + \gB^{ } f^{abc}_{ } T^c_{ }\, \Tint{R,S} 
%  \!\! A^a_i (R) A^b_\mu (S)\; \deltabar(R+S-Q)
%  \;. 
% \ee

For calibration, we may once again start with LO results. 
For the denominator, \eqs\nr{partition_nr_lo} and \nr{N_0_HQET_lo}
continue to hold, i.e.\ 
$\mathcal{N}_0^\rmii{IR} = 2  + \rmO(\gB^2)$. 
For the numerator, 
extracting external states like between 
\eqs\nr{partition_nr_lo} and \nr{N_0_HQET_lo}, 
the amplitude from \eq\nr{numerator_def3}
evaluates to 
\be
 \mathcal{F}^\rmii{IR}_i 
  = 
   2 \gB^{ }  
 \bigl[ 
   \EA % - \EB
 + \BA % - \BB
 \bigr] 
 + \rmO(\gB^3)
 \;. \la{F_i_HQET_lo}
\ee
The ratio of 
\eq\nr{F_i_HQET_lo} and $\mathcal{N}_0^\rmii{IR}$ 
yields a Lorentz force like in \eq\nr{lorentz_em}.

Proceeding to NLO, let us start by 
elaborating on the issue of the mass counterterm, which previously played 
an important role in cancelling singular propagators $\sim \rmO(1/Q)$.
The counterterm takes a form obtained from the $M\to\infty$ limit
of \eq\nr{deltaM_NR}, {\it viz.} 
\be 
 \delta M = - \gB^2\CF \Tint{R}
 \frac{1}{ R^2 \Lambda^{ }_{P-R}}
 \; + \; \rmO(\gB^4)
 \;. \la{deltaM_IR}
\ee
The Matsubara sum can be extracted from \eq\nr{sum1}.
At the on-shell point, $\omega^{ }_n\to i M$, and omitting exponentially
small terms $\sim e^{-M/T}$, this yields 
\be 
 \delta M = - \gB^2\CF \int_{\vec{r}} \frac{1}{2\epsilon_r^2}
 \; + \; \rmO(\gB^4)
 \;. \la{deltaM_IR_2}
\ee
Recalling that after resummation 
the temporal gauge field components, 
which are responsible for \eq\nr{deltaM_IR}, 
carry a thermal mass $\mD^{ }$,
\eq\nr{deltaM_IR} corresponds to a well-known  
correction to a heavy-quark mass, 
$M |^{ }_\T \supset -g^2\CF \mD^{ }/(8\pi)$~\cite{htl1}. 
However, as explained above,
we do not need to carry out resummation 
in our actual computation.
Therefore the mass counterterm gives no contribution in 
dimensional regularization. 

With this framework, the denominator remains at the
value of \eq\nr{N_0_HQET_nlo}, {\it viz.} 
$
 \mathcal{N}^\rmii{IR}_0 = 
 \mathcal{N}^\rmii{HQET}_0 
$.
The numerator is determined by the graphs in \fig\ref{fig:IR}. 
Many terms proportional to~$\CF$ vanish, 
for the same reason that the mass counterterm does not contribute. 
The gluon self-energy can 
be set aside, as it agrees with that on the high-energy side
and therefore drops out in the matching. The other diagrams 
on the second row of \fig\ref{fig:IR} produce a non-vanishing 
contribution proportional to $\CA$. 

The sum-integrals obtained after a Taylor expansion are in the same
class as those discussed in \se\ref{se:nonrel}.  
Writing $Z^{ }_{\!\E,\B} = 1+ \delta Z^{ }_{\!\E,\B}$
where $\delta Z^{ }_{\!\E,\B}\sim\rmO(\gB^2)$,
we are left with 
\ba
% && \hspace*{-1.0cm} 
 \frac{ \mathcal{F}^\rmii{IR}_i  }
     { 2 } 
 & = & 
 \gB^{ } 
      \bigl[  
        \EA
       + 
        \BA
       % -
       %  \BB
      \bigr]
  \biggl\{
  1\,    -\,
 \gB^2 \CF \Tint{R} 
      \biggl[
       \frac{1}{R^2\Lambda^2_{P-R}}
       + 
       \frac{1 - \xi}{R^4} 
      \biggr]
 \biggr\}
 \nn[2mm] 
%%%%%%%
 & + & \frac{ \gB^3 \CA }{2} \Tint{R} \biggl\{ 
      \biggl[
       \frac{1}{R^2\Lambda^2_{P-R}}
       + 
       \frac{(d-3)(\xi - 1) - 2}{R^4} 
      \biggr] \EA
% \nn[2mm] 
%%%%%%%
% & & \hspace*{1.3cm}
     \, + \,
      \biggl[
       \frac{1}{R^2\Lambda^2_{P-R}}
       -
       \frac{4}{R^4} 
      \biggr]
      % \Bigl[
          \BA
      % - \BB \Bigr]
      \biggr\}
 \nn[2mm] 
%%%%%%%
 & + & 
 \gB^{ }\, \EA \, \delta Z^{ }_{\!\E} 
 \; + \; \gB^{ }\,
      % \Bigl[  
        \BA
      % -
      %  \BB
      %\Bigr]
      \, \delta Z^{ }_\B
% \nn[2mm] 
%%%%%%%%
% & + &
 \, + \, 
 \mbox{(gluon self-energy)}
 \; + \; \rmO(\gB^5) 
 \;. \hspace*{6mm}
 \la{F_i_IR_nlo}
\ea
The correction 
proportional to $\CF$ agrees with that 
in \eq\nr{N_0_HQET_nlo}, and therefore drops out in 
the ratio of \eq\nr{matching}.  
Inserting the values of the master sum-integrals
from \eqs\nr{master_sum1}, \nr{master_sum2} 
and \nr{master_sum6}, finally yields
\ba
%%%%%%%
 \frac{ \mathcal{F}^\rmiii{IR}_i  }
      { \mathcal{N}^\rmiii{IR}_0  }      
 & = &  
 \gB^{ }\, \EA \, 
 \biggl\{ 
 1 
 \; + \; 
 \delta Z^{ }_{\!\E}  
 \; + \; 
 \frac{ \gB^2 \CA (3 - \xi) }{(4\pi)^2} 
 \biggr\} 
 \nn[0mm]
%%%%%%%%
 & + & 
 \gB^{ }\,
     % \Bigl[  
        \BA
     %  -
     %   \BB
     % \Bigr]
     \,
  \biggl\{ 
 1 
 \; + \;
 \delta Z^{ }_{\!\B}
 \; - \; 
 \frac{ \gB^2 \CA \,\mu^{-2\epsilon} }{(4\pi)^2} 
 \biggl[ 
       \frac{1}{\epsilon} 
   + 2 \ln\biggl( \frac{\bmu e^{\gammaE}}{4\pi T} \biggr)
   - 2 
 \biggr] 
 \biggr\}
 \nn[3mm] 
%%%%%%%
 & + & \mbox{(gluon self-energy)}
 \; + \; \rmO\bigl( \epsilon\,\gB^3 ,\gB^5 \bigr) 
 \;. \la{a_i_ir_nlo_expl}
\ea

%%%%%%%%%%%%%%%%%%%%%%%%%%% SECTION %%%%%%%%%%%%%%%%%%%%%%%%%%%%%%%%%%%%%%
%
\section{Result and discussion}
\la{se:discussion}

In the preceding sections we have computed the objects 
appearing in \eq\nr{matching} in three different ways: 
in vacuum, whereby would-be loop effects originate from 
the scale $\sim M$ but all cancel 
in the end (cf.\ \se\ref{se:qcd_vacuum}); 
at finite temperature, by making use of a Noether current
and its time derivative, thereby 
incorporating effects from the thermal scales $\sim \sqrt{MT}, T$
(cf.\ \ses\ref{se:qcd_thermal} and \ref{se:nonrel});
and in an IR description, which makes use of a Lorentz 
force operator rather than a Noether current
(cf.\ \se\ref{se:ir}). By requiring that the results agree, 
we can determine the 
renormalization constants of the Lorentz force operator, 
defined through \eq\nr{F_IR}. 
Concretely, a comparison of 
\eqs\nr{A_i_nr_nlo_expl} and \nr{a_i_ir_nlo_expl} yields 
\ba
 Z^{ }_{\!\E} & = & 
 1 + \delta Z^{ }_{\!\E} \; = \; 
 1 + \rmO(g^4)
 \;, \la{dZE} \\ 
 Z^{ }_{\!\B} & = & 
 1 +  \delta Z^{ }_{\!\B} \; = \; 
 1 + \frac{ g^2 \CA }{(4\pi)^2} 
 \biggl[ 
       \frac{1}{\epsilon} 
   + 2 \ln\biggl( \frac{\bmu e^{\gammaE}}{4\pi T} \biggr)
   - 2 
 \biggr] 
 + \rmO(g^4) 
 \;, \la{dZB}
\ea
where we have replaced the bare coupling $\gB^2$ 
by its renormalized value, {\it viz.}
\be
 \gB^2\, \mu^{-2\epsilon} = g^2 + \frac{g^4 }{(4\pi)^2} 
 \frac{2\Nf - 11\Nc}{3\epsilon} + \rmO(g^6)
 \;. \la{gB}
\ee

It is a little bit subtle to see
which scales have been integrated out
through the matching steps that we have presented. 
Indeed, even though $T$
appears inside the logarithm
in \eq\nr{dZB}, it has not been fully eliminated, 
but still affects the low-energy observables that could be measured
with \eq\nr{F_IR}, such as \eq\nr{GB_def}. Rather, what has been eliminated
are the heavy quark spatial momenta. These appear 
explicitly in the HQET Noether current, $J^\rmii{HQET}_i$, 
which contains derivatives acting on $\theta^\dagger$ and $\theta$, 
but are absent from \eq\nr{F_IR}, where $\vec{v}$ appears
as an external parameter. In a 2-point
correlator, the velocity appears 
in the form~$
 \langle \vec{v}^2 \rangle^{ }_{ }
$.
It has been pointed out in ref.~\cite{kappaE} that a field-theoretic
interpretation for this average is given by the $\tau$-independent 
part of the 2-point imaginary-time 
correlator of the vector current,\footnote{%
 Or, in real frequency space, by the area under the transport peak
 in the corresponding spectral function.
 } 
$\int_\vec{x} J_i^\rmii{QCD}$, 
normalized to the susceptibility. This quantity has been computed
up to NLO in \eqs(3.4), (3.5), (4.1) and (4.5) of ref.~\cite{GVtau}.
It accounts for the dynamics at the momentum scale $p\sim\sqrt{MT}$,  
and is finite after mass renormalization, indicating that this 
physics does not mix with the renormalization of the magnetic field
at this order. 
The present computation 
has thus accounted for thermal gauge modes kicking the heavy quarks
in spatial directions, and left over are thermal gauge modes 
not involved in such momentum transfer.  

Given the subtle interpretation, it is comforting that 
the $1/\epsilon$-parts of \eqs\nr{dZE} and \nr{dZB} can be compared
with literature. In the thermal context 
one considers 2-point correlators of the Lorentz
force, normalized to the 2-point correlator of
the Noether charge (i.e.\ susceptibility). 
For the magnetic field this leads to~\cite{1overM} 
\be
 G^{ }_{\!\B}(\tau) \; \equiv \;
 \frac{ \gB^2 
        \sum_i \re\tr\langle U(\beta;\tau)\, B^{ }_i(\tau)\, 
                      U(\tau;0) \, B^{ }_i(0)\, \rangle
      }{
        3 \re\tr\langle U(\beta;0) \rangle
      }
 \;, \la{GB_def}
\ee
where $U$ is a timelike Wilson line and the trace is now
in colour space. 
The imaginary-time correlator
is conveniently viewed in a spectral representation, 
\be
 G^{ }_{\!\B}(\tau)
 = 
 \int_0^\infty \! \frac{{\rm d}\omega}{\pi}
 \, \rho^{ }_{\!\B}(\omega) \, 
 \frac{\cosh [\omega (\frac{\beta }{2}-\tau) ] }
      {\sinh [ \frac{\omega \beta }{2} ] }
 \;. \la{spectral}
\ee
For the electric counterpart, a general argument~\cite{kappaE} as well as 
a 1-loop computation~\cite{rhoE} show that the spectral
function $\rho^{ }_{\!\E}$
is rendered finite through gauge coupling renormalization, 
and this is consistent with \eq\nr{dZE}.\footnote{%
 In lattice regularization, 
 a finite renormalization factor
 of $\rmO(g^2)$ is however needed~\cite{renormE}.
 } 
In contrast, for $G^{ }_{\!\B}$, a 1-loop 
computation~\cite{1overM} shows that after gauge coupling 
renormalization, 
the spectral function is {\em not finite}, but rather reads
\be
 \rho^{ }_{\!\B}(\omega) 
  = 
  \frac{g^2\CF^{ }\omega^3}{6\pi}
  \, 
  \biggl[
   1 - \frac{g^2\CA}{(4\pi)^2} \frac{2}{\epsilon} + \mbox{(finite)}
  \biggr] 
 + \rmO(g^6)
 \;. \la{rhoB}
\ee
We now see from \eq\nr{dZB} that multiplying the magnetic fields
by $Z^{ }_{\!\B}$, 
i.e.\ considering
the correlator 
$
 Z^{2}_{\!\B} \, 
 G^{ }_{\!\B}(\tau)
$, 
the divergence in \eq\nr{rhoB} duly cancels. 

A completely different crosscheck originates from vacuum 
computations, concerning the operator multiplied by $c^{ }_\B$ in 
\eq\nr{S_nr}, known as the chromomagnetic moment. In our notation, 
the 1-loop result for $c^{ }_\B$~\cite{hqet1} can be 
expressed as 
\be
 c^{ }_\B = 1 + \frac{g^2}{(4\pi)^2}
 \biggl\{ 
    \CA^{ } \biggl[ \frac{1}{\epsilon} + \ln\frac{\bmu^2}{M^2} +2  \biggr] 
  + 2 \CF^{ }
 \biggr\} 
 + \rmO(g^4)
 \;. \la{cB}
\ee
Even if the chromomagnetic moment concerns spin-dependent effects, 
the magnetic field appears in the same form
in \eqs\nr{S_nr} and \nr{F_IR},  $\sim \gB^{ }\vec{B}$. 
Indeed the anomalous dimension visible in
\eq\nr{cB} agrees with that in \eq\nr{dZB}. 

Going to higher orders, we could possibly 
profit from the fact that the anomalous dimension of 
the chromomagnetic moment has 
been determined up to 2-loop~\cite{hqet2a,hqet2b} and 
3-loop level~\cite{hqet3}. 
Furthermore, non-perturbative renormalization
in terms of a renormalization group invariant (RGI) operator has
been worked out~\cite{renormB}, corresponding to 
$\bmu\to\infty$. After such a non-perturbative
renormalization, results should be run down to the $\msbar$  scale 
$\bmu \simeq 4\pi T e^{1 - \gammaE} \approx 19.179 T$
according to \eq\nr{dZB}. Given that no pole mass ambiguity appears, 
unlike in \eq\nr{cB}, 
and that there is a large numerical prefactor, 
a reasonable precision could be hoped for.   

To summarize, 
all ingredients needed for estimating the influence 
of magnetic interactions on heavy quark diffusion should now
be available, at least in an approximate form. 

%%%%%%%%%%%%%%%%%%%%%%%%%%% SECTION %%%%%%%%%%%%%%%%%%%%%%%%%%%%%%%%%%
%
\section*{Acknowledgements}

M.L.\ thanks Debasish Banerjee
and Saumen Datta for helpful discussions. 
This work was supported by the Swiss National Science Foundation
(SNF) under grant 200020B-188712.

\small{
%%%%%%%%%%%%%%%%%%%%%%%%%% BIBLIO %%%%%%%%%%%%%%%%%%%%%%%%%%%%%%%%%%%%%%%%%
%

}

\end{document}